\documentclass[
  reprint,
  aps,
  prx,
  amsmath,amssymb,
  superscriptaddress,
  longbibliography,
  nofootinbib,
  floatfix
]{revtex4-2}
\usepackage[utf8]{inputenc}
\usepackage{graphicx}
\usepackage{dcolumn}
\usepackage{bm}
\usepackage{tikz}
\usepackage{quantikz}
\usepackage[caption=false]{subfig}
\usepackage{microtype}
\usepackage{float}
\usepackage[colorlinks=true, linkcolor=blue, citecolor=blue, urlcolor=blue]{hyperref}

\begin{document}
\title{Quantum Parrondo's Paradox via a Single Phase Defect with Symmetry Breaking and Directed Transport}
\author{Jen-Yu Chang}
\email{leo07010@gmail.com}
\affiliation{Department of Electrophysics, National Yang Ming Chiao Tung University, Hsinchu, Taiwan}
\author{Yun-Hsuan Chen}
\affiliation{Quantum Information Center, Chung Yuan Christian University, Taoyuan, Taiwan}
\author{Gooi Zi Liang}
\affiliation{Quantum Information Center, Chung Yuan Christian University, Taoyuan, Taiwan}

\author{Chih-Yu Chen}
\affiliation{Quantum Information Center, Chung Yuan Christian University, Taoyuan, Taiwan}
\affiliation{Undergraduate Program in Intelligent  Computing and Big Data, Chung Yuan Christian University, Taoyuan, Taiwan}

\author{Tsung-Wei Huang}
\affiliation{Quantum Information Center, Chung Yuan Christian University, Taoyuan, Taiwan}
\affiliation{Master Program in Intelligent Computing and Big Data, Chung Yuan Christian University, Taoyuan, Taiwan}

\date{\today}

\begin{abstract}
Parrondo's paradox describes the counterintuitive phenomenon in which alternating two individually losing games yields a winning outcome. Extending this effect to the quantum regime has typically required high-dimensional coin spaces, entangled initial states, or engineered decoherence. Here we show that a genuine and persistent quantum Parrondo effect can be realized with minimal resources---a single-qubit coin, a fixed periodic sequence of two $SU(2)$ operators, and a single localized phase defect $e^{i\phi}$ at the origin of a discrete-time quantum walk. By breaking translational symmetry, the phase defect acts as a scattering center that enables momentum mixing and interference-induced rectification, converting two losing games into a directed ``quantum ratchet.'' We critically reassess the winning criterion and demonstrate that the position expectation value $\langle \hat{x} \rangle$, rather than the commonly used probability asymmetry $P_R - P_L$, is the appropriate metric for validating the paradox. Harmonic analysis of the drift velocity reveals a complex, resonance-type dependence on $\phi$ with high-order Fourier components, reflecting nontrivial multi-path interference at the defect site. We further show that winning strategies are associated with cyclic restoration of coin-position entanglement, and that the ratchet effect is robust across a wide range of initial states. Our results establish that spatial inhomogeneity, rather than additional quantum resources, is the essential ingredient for a sustainable quantum Parrondo effect, offering a resource-efficient blueprint for directed transport on near-term quantum platforms.
\end{abstract}
\maketitle
\section{Introduction}
Quantum walks, the quantum mechanical counterparts of classical random walks, have emerged as a foundational framework in quantum information science~\cite{venegas2012quantum}. Since their inception by Aharonov et al.~\cite{aharonov1993quantum} and subsequent formalization by Kempe~\cite{kempe2003quantum}, they have proven invaluable for algorithm design and quantum simulation. Leveraging quantum superposition and interference, quantum walks exhibit a ballistic spreading behavior that offers a quadratic speed-up over the diffusive spreading of classical random walks~\cite{ambainis2003quantum}. This unique property underpins a diverse array of applications, including efficient quantum search algorithms~\cite{shenvi2003quantum}, universal quantum computation~\cite{childs2013universal}, models of quantum transport in complex systems~\cite{perets2008realization}, and the investigation of topological phases and localization in disordered media~\cite{kitagawa2010exploring,Kitagawa2012}. Experimentally, quantum walks have been realized across a wide range of platforms, from trapped ions and neutral atoms~\cite{zahringer2010realization,karski2009quantum,schmitz2009quantum} to integrated photonic chips~\cite{tang2018experimental,xue2015experimental} and programmable superconducting processors~\cite{yan2019strongly,gong2021quantum,flurin2017observing}, demonstrating the maturity and versatility of the framework.

A particularly intriguing phenomenon within the realm of stochastic processes is Parrondo's paradox. Originally formulated by Parrondo and systematized by Harmer and Abbott~\cite{harmer1999parrondo,harmer1999losing}, the paradox describes a counterintuitive scenario where alternating between two individually losing games results in a winning strategy. In its classical form, the paradox relies on a combination of biased coin flips (Game~A) and capital-dependent rules (Game~B) to rectify Brownian motion, a mechanism intimately linked to the physics of Brownian ratchets~\cite{chen2010quantum,reimann2002brownian}. Despite the individual negativity of each game, specific periodic sequences, such as AB or ABB, can yield a positive expected gain~\cite{parrondo1996how,harmer2000parrondos,dinis2003optimal}. Beyond its theoretical interest, the paradox has found interdisciplinary applications in genetics~\cite{Reed2007Parrondo}, population dynamics and ecology~\cite{cheong2017nomadic,tan2021alternating}, financial modeling~\cite{stutzer2003simple,Spurgin01032005}, and the design of Brownian thermal engines~\cite{parrondo2003paradoxical}, underscoring the broad relevance of the counterintuitive insight that losing strategies can combine to win.

The extension of Parrondo's paradox to the quantum domain has garnered significant interest, typically realized through discrete-time quantum walks (DTQWs) in which classical coin flips are replaced by unitary operators acting on a quantum coin space, usually $SU(2)$ rotations~\cite{meyer1996from}. Early theoretical work by Flitney and Abbott~\cite{flitney2002quantum} and Meyer~\cite{meyer2002quantum} demonstrated that quantum interference could indeed produce Parrondo-like effects~\cite{flitney2012quantum}. However, subsequent analysis by Flitney, Ng, and Abbott~\cite{flitney2002games} revealed a critical limitation: in simple homogeneous quantum walks, these effects are often transient, decaying to zero in the asymptotic limit. This led to a prevailing assumption that single-qubit coin systems were insufficient for sustaining a robust quantum Parrondo effect without additional resources.

To overcome this limitation, a wide variety of extensions to the standard DTQW framework have been proposed. These include history-dependent coin schemes~\cite{bleiler2011properly}, higher-dimensional coin spaces such as qutrits~\cite{rajendran2018playing}, two-coin protocols~\cite{rajendran2018implementing}, open quantum systems with decoherence~\cite{ding2012quantum}, engineered noise~\cite{romanelli2009distribution,walczak2023noise}, time-dependent coin operators~\cite{Pires2020time}, chaotic switching sequences~\cite{Lai2021chaotic}, deterministic aperiodic coin sequences~\cite{Walczak2021aperiodic}, different shift operators~\cite{walczak2024parrondo_shift,shikano2010localization}, and entangled initial states~\cite{rajendran2010parrondo}. A common thread among these proposals is the introduction of additional complexity or resources to break the symmetry that suppresses the paradox. More recently, the Parrondo effect has also been extended to continuous-time quantum walks~\cite{ximenes2024parrondo} and to quantum coin-toss simulations~\cite{Lai2020random}, further broadening the scope of the phenomenon.

Despite this extensive body of work, the fundamental criteria for a winning quantum game remain subjects of intense debate. While \citet{chandrashekar2011parrondo} proposed that specific sequences of $SU(2)$ operations could yield observable Parrondo-like effects using a single-qubit coin, and subsequent experimental work by \citet{jan2020experimental} confirmed a persistent positive drift on an optical platform, these conclusions are primarily based on observing shifts in the probability distribution. We argue that defining a winning strategy through simple probability heuristics fails to capture the mathematical and economic essence of Parrondo's paradox: the emergence of a positive \textbf{expected gain} from individually losing games.

Recent progress has opened a new direction for resolving these issues. \citet{PhysRevA.110.052440} demonstrated that the paradox can be robustly realized using \textit{inhomogeneous coins} without the need for high-dimensional coin spaces or complex temporal switching, while \citet{kadiri2024scouring} provided a systematic framework for identifying paradoxical regimes in DTQWs. Concurrently, \citet{walczak2025parrondo} showed that Parrondo's paradox can emerge in space-inhomogeneous quantum walks and that the effect can nontrivially influence the entanglement dynamics. Inspired by these developments and the realization of maximal coin-walker entanglement in inhomogeneous ballistic walks~\citep{PhysRevA.105.042216}, we introduce a \textbf{position-dependent phase modification}---specifically a localized phase shift $e^{i\phi}$ at the origin ($x=0$)---as a minimal form of spatial inhomogeneity. Unlike previous approaches that rely on complex temporal switching protocols or multi-coin architectures, our model retains a simple periodic game sequence with a single-qubit coin. We provide an analytical proof demonstrating that in the absence of this phase defect, the homogeneous system is confined to ballistic transport governed by momentum conservation, rendering a true ratchet effect impossible. The introduction of the local phase breaks momentum conservation, acting as a scattering center that facilitates momentum mixing and interference-induced rectification.

A central contribution of our work is the critical reassessment of the winning criterion for quantum Parrondo games. We demonstrate that the position expectation value $\langle \hat{x} \rangle$, rather than the probability asymmetry $P_R - P_L$ employed in previous studies~\cite{chandrashekar2011parrondo,jan2020experimental}, is the appropriate metric for validating the paradox in a physical and economic sense. Using this rigorous criterion, we show that while homogeneous periodic sequences may appear to ``win'' via parameter-tuned ballistic transport, a genuine quantum Parrondo ratchet---one that yields a net positive expected value from losing components---is only sustainable by breaking translational symmetry. We validate this framework through systematic numerical simulations, harmonic analysis of the drift velocity, investigation of entanglement dynamics, and verification of robustness against initial state variations. The present model constitutes a minimal realization of a genuine quantum Parrondo ratchet: it employs a single-qubit coin, unitary dynamics, a fixed periodic sequence, and a single localized phase defect, without entanglement, decoherence, higher-dimensional coins, or time-dependent control.

The remainder of this paper is organized as follows. Section~\ref{sec:methods} introduces the theoretical framework, including the classical paradox, the DTQW formalism, the phase-defect model, and the analytical symmetry-breaking argument. Section~\ref{sec:results} presents the numerical results: the demonstration of the paradox under general and standard coin parameters (Sec.~\ref{sec:numerical_results}), the reassessment of the winning criterion (Sec.~\ref{sec:winning_criterion}), the phase-controlled transport analysis (Sec.~\ref{sec:phase_controlled}), the harmonic decomposition of the drift velocity (Sec.~\ref{sec:harmonic_analysis}), the entanglement dynamics (Sec.~\ref{sec:entanglement}), and the initial state independence (Sec.~\ref{sec:initial_state}). Section~\ref{sec:conclusion} summarizes our findings and discusses future directions.
\section{Framework and Symmetry Breaking}
\label{sec:methods}
\subsection{Classical Parrondo's paradox}
\label{sec:classical_parrondo}
\begin{figure}[h]
    \centering
    \includegraphics[width=\linewidth]{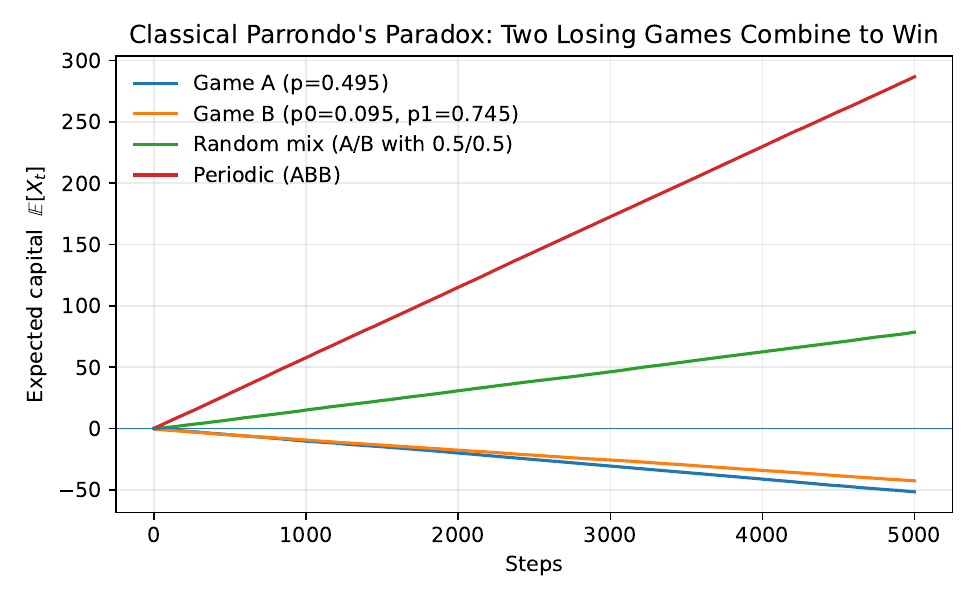}
    \caption{\textbf{Demonstration of the classical Parrondo's paradox.} The evolution of expected capital $\mathbb{E}[X_t]$ shows that while individual Game~A (blue) and Game~B (orange) are losing strategies, switching between them randomly (green) or in a periodic $ABB$ sequence (red) results in a winning trend.}
    \label{fig:parrondo_classical}
\end{figure}

Parrondo's paradox describes the counterintuitive emergence of a winning expectation from the combination of individually losing dynamics~\cite{harmer1999parrondo,parrondo1996how}. Consider two games, $G_A$ and $G_B$, each characterized by a negative long-time drift of the player's capital:
\begin{equation}
v_A \equiv \lim_{t\to\infty}\frac{\mathbb{E}[X_t^{(A)}]}{t} < 0, \qquad
v_B \equiv \lim_{t\to\infty}\frac{\mathbb{E}[X_t^{(B)}]}{t} < 0,
\label{eq:losing_games}
\end{equation}
where $X_t^{(i)}$ denotes the capital after $t$ rounds of the game $G_i$. Therefore,$\mathbb{E}$ means the expected value after the $t$ round of the game and $v$ the average expected value for each round. In the following discussion, the capital, $X$, is defined on a one dimensional number line, and then the $v$ is the average velocity of the capital or called drift. The paradox is realized when a combined strategy---either a deterministic periodic sequence (e.g., $ABB$) or a stochastic mixture---yields a positive drift:
\begin{equation}
v_{AB} > 0.
\label{eq:winning_combined}
\end{equation}
In the canonical classical formulation~\cite{harmer2000parrondos}, Game~A is a simple biased coin flip with a slight losing probability $p_A = 1/2 - \epsilon$, while Game~B employs a capital-dependent rule: the coin bias switches depending on whether the current capital is divisible by a modulus $M$ (typically $M=3$). Neither game is favorable on its own, yet their alternation exploits the capital-dependent switching to produce a net rectification of the random walk, in close analogy with Brownian ratchets~\cite{reimann2002brownian,parrondo2003paradoxical}. Figure~\ref{fig:parrondo_classical} illustrates a typical realization: both individual games lose on average, while the periodic sequence $ABB$ and random switching both yield a growing expected capital.
Two features of the classical paradox are essential for its quantum generalization. First, the \emph{winning criterion} is defined through the expected capital $\mathbb{E}[X_t]$---not through a probability asymmetry such as $\Pr(X_t>0)$---since the latter can be misleading when the distribution has broad tails. Second, the rectification mechanism requires a form of \emph{state-dependent asymmetry} (the capital-dependent rule in Game~B) that breaks the detailed balance of the individual games. In the quantum setting, we will show that these roles are played by the position expectation value $\langle \hat{x} \rangle$ and by a localized phase defect that breaks the translational symmetry of the lattice, respectively.
\subsection{Discrete-time quantum walks}
\label{sec:dtqw}
We consider a discrete-time coined quantum walk (DTQW) on the one-dimensional integer lattice $\mathbb{Z}$~\cite{aharonov1993quantum,kempe2003quantum}. The total Hilbert space is
\begin{equation}
\mathcal{H} = \mathcal{H}_P \otimes \mathcal{H}_C,
\label{eq:hilbert}
\end{equation}
where $\mathcal{H}_P = \mathrm{span}\{\ket{x} : x \in \mathbb{Z}\}$ is the position (walker) space and $\mathcal{H}_C = \mathrm{span}\{\ket{0},\ket{1}\}$ is the two-dimensional coin space. We adopt the convention that position indices appear on the left throughout.
A single time step consists of a coin rotation followed by a conditional shift:
\begin{equation}
\hat{U} = \hat{S}\,\big(\hat{I}_P \otimes \hat{C}\big),
\label{eq:U_def}
\end{equation}
where $\hat{C}$ is a (possibly position-dependent) unitary coin operator acting on $\mathcal{H}_C$. The conditional shift operator translates the walker left or right depending on the coin state:
\begin{equation}
\hat{S} = \sum_{x\in\mathbb{Z}} \Big(
  \ket{x+1}\!\bra{x} \otimes \ket{0}\!\bra{0}
+ \ket{x-1}\!\bra{x} \otimes \ket{1}\!\bra{1}
\Big).
\label{eq:S_def}
\end{equation}
The coin operator is parametrized as a general $SU(2)$ rotation~\cite{meyer1996from}:
\begin{equation}
\hat{C}(\alpha,\beta,\gamma) =
\begin{pmatrix}
e^{i\alpha}\cos\beta  & -e^{-i\gamma}\sin\beta \\
e^{i\gamma}\sin\beta  &  e^{-i\alpha}\cos\beta
\end{pmatrix},
\label{eq:SU2}
\end{equation}
where $\alpha,\gamma \in [0,2\pi)$ are phase parameters and $\beta \in [0,\pi]$ controls the mixing angle between the two coin states. Throughout this work, we denote by $\hat{C}_A \equiv \hat{C}(\alpha_A,\beta_A,\gamma_A)$ and $\hat{C}_B \equiv \hat{C}(\alpha_B,\beta_B,\gamma_B)$ the coin operators for Games~A and~B, respectively.
In the quantum Parrondo framework, the walker's position plays the role of capital, and we define the \emph{game outcome} through the asymptotic drift velocity of the position expectation value:
\begin{equation}
v \equiv \lim_{T\to\infty}\frac{\mathbb{E}(x)}{T}=\lim_{T\to\infty} \frac{\langle \hat{x} \rangle_T}{T},
\label{eq:drift_def}
\end{equation}
where $\langle \hat{x} \rangle_T = \bra{\Psi(T)} \hat{x} \ket{\Psi(T)}$ with $\hat{x} = \sum_x x\,\ket{x}\!\bra{x} \otimes \hat{I}_C$. A game is \emph{losing} if $v < 0$ and \emph{winning} if $v > 0$. The paradox is realized when $v_A < 0$, $v_B < 0$, but $v_{AB} > 0$ for a combined sequence.
\subsection{Homogeneous walks and momentum-space reduction}
\label{sec:methods_homo}
A homogeneous walk is one in which the coin operator is position-independent, i.e., the total coin acts as $\hat{I}_P \otimes \hat{C}_0$ for a fixed $\hat{C}_0 \in SU(2)$. The resulting step operator is $\hat{U}_{\mathrm{homo}} = \hat{S}(\hat{I}_P \otimes \hat{C}_0)$. We now show that $\hat{U}_{\mathrm{homo}}$ commutes with the lattice translation operator $\hat{T}_d \equiv \hat{T}_d^{(P)} \otimes \hat{I}_C$, where $\hat{T}_d^{(P)}\ket{x} = \ket{x+d}$.
\emph{Proof.}---Since $\hat{I}_P \otimes \hat{C}_0$ is manifestly translation-invariant ($[\hat{T}_d,\,\hat{I}_P \otimes \hat{C}_0] = 0$), it suffices to show that $\hat{S}$ commutes with $\hat{T}_d$. Acting on an arbitrary basis state $\ket{x}\otimes\ket{c}$ with $c \in \{0,1\}$:
\begin{align}
\hat{S}\,\hat{T}_d\,\ket{x}\otimes\ket{c}
&= \hat{S}\,\ket{x+d}\otimes\ket{c}
= \ket{x+d+s_c}\otimes\ket{c}, \nonumber \\
\hat{T}_d\,\hat{S}\,\ket{x}\otimes\ket{c}
&= \hat{T}_d\,\ket{x+s_c}\otimes\ket{c}
= \ket{x+s_c+d}\otimes\ket{c},
\label{eq:ST_commute_proof}
\end{align}
where $s_0 = +1$ and $s_1 = -1$ are the shift displacements defined in Eq.~\eqref{eq:S_def}. Since the two expressions coincide for all $x$, $c$, and $d$, we have $[\hat{S},\hat{T}_d] = 0$ and therefore
\begin{equation}
[\hat{U}_{\mathrm{homo}},\,\hat{T}_d] = 0 \qquad \forall\, d \in \mathbb{Z}.
\label{eq:commute_condition}
\end{equation}
Physically, this states that a uniform lattice with a position-independent coin has no preferred origin; the dynamics are identical at every site.
This translational invariance permits block diagonalization in quasi-momentum space. Introducing the Fourier basis
\begin{equation}
\ket{k} = \frac{1}{\sqrt{2\pi}} \sum_{x\in\mathbb{Z}} e^{ikx}\ket{x}, \qquad k \in (-\pi,\pi],
\label{eq:k_basis}
\end{equation}
the shift operator reduces to the $2\times 2$ coin-space matrix
\begin{equation}
\bm{\Sigma}(k) =
\begin{pmatrix}
e^{-ik} & 0 \\
0        & e^{ik}
\end{pmatrix},
\label{eq:S_k}
\end{equation}
and the one-step unitary likewise becomes a $2\times 2$ matrix:
\begin{equation}
\bm{\mathcal{U}}_{\mathrm{homo}}(k) = \bm{\Sigma}(k)\,\hat{C}_0.
\label{eq:U_k}
\end{equation}
Note that the coin operator $\hat{C}_0$ acts solely on the two-dimensional coin space $\mathcal{H}_C$ and is therefore already a $2\times 2$ matrix; no further projection is needed. Throughout this paper, we use boldface Greek letters to denote the $k$-space (momentum-space) $2\times 2$ representations of operators that originally involve position space, reserving hatted Latin letters ($\hat{\cdot}$) for the full-Hilbert-space operators. Specifically: $\hat{S} \to \bm{\Sigma}(k)$ for the shift, and $\hat{U} \to \bm{\mathcal{U}}(k)$ for the step operator. Coin-space operators such as $\hat{C}_0$ are already $2\times 2$ matrices and retain their original notation.
Let $\lambda_\pm(k) = e^{-i\omega_\pm(k)}$ denote the eigenvalues of $\bm{\mathcal{U}}_{\mathrm{homo}}(k)$, with corresponding eigenstates $\ket{u_k^\pm}$. These define the quasi-energy bands $\omega_\pm(k)$ and group velocities
\begin{equation}
v_g^{(\pm)}(k) = \frac{d\omega_\pm(k)}{dk}.
\label{eq:group_vel}
\end{equation}
For a general initial state $\ket{\Psi(0)} = \ket{x\!=\!0} \otimes \ket{\chi_c}$ localized at the origin, its position-space component in the Fourier basis~\eqref{eq:k_basis} is $\braket{k|x\!=\!0} = 1/\sqrt{2\pi}$, so every quasi-momentum mode is populated with equal weight: $\langle k|\Psi(0)\rangle_P = \ket{\chi_c}/\sqrt{2\pi}$. The asymptotic drift velocity of the homogeneous walk is then given by
\begin{equation}
v_{\mathrm{homo}}
= \int_{-\pi}^{\pi}\! dk \sum_{\sigma=\pm} v_g^{(\sigma)}(k)\,
\big|\!\braket{u_k^\sigma|\chi_c}\!\big|^2,
\label{eq:vhomo}
\end{equation}
where the overlap $|\!\braket{u_k^\sigma|\chi_c}\!|^2$ determines how the initial coin state projects onto each band.
To implement Parrondo protocols, we employ periodic sequences of coin operators. For a period-$p$ sequence $s_1 s_2 \cdots s_p$ with $s_j \in \{A,B\}$ (e.g., $ABB$ has $p=3$), the single-period (Floquet) unitary is
\begin{equation}
\hat{U}_{\mathrm{seq}} = \hat{U}_{s_p} \hat{U}_{s_{p-1}} \cdots \hat{U}_{s_1},
\qquad \hat{U}_{s_j} = \hat{S}\big(\hat{I}_P \otimes \hat{C}_{s_j}\big).
\label{eq:Useq_def}
\end{equation}
Because each factor $\hat{U}_{s_j}$ is homogeneous, $\hat{U}_{\mathrm{seq}}$ inherits translational invariance and remains block-diagonal in $k$:
\begin{equation}
\bm{\mathcal{U}}_{\mathrm{seq}}(k) = \prod_{j=p}^{1} \big[\bm{\Sigma}(k)\,\hat{C}_{s_j}\big]
\equiv \bm{\mathcal{U}}_{\mathrm{eff}}(k).
\label{eq:Useq_k}
\end{equation}
This defines a stroboscopic effective $2\times 2$ Floquet operator $\bm{\mathcal{U}}_{\mathrm{eff}}(k)$ whose band structure fully determines the drift. Crucially, because quasi-momentum $k$ remains a good quantum number, any drift in a homogeneous periodic walk is entirely determined by the band dispersion of $\bm{\mathcal{U}}_{\mathrm{eff}}(k)$---that is, by parameter mixing of coin angles---rather than by a position-dependent ratchet mechanism. As we demonstrate numerically in Sec.~\ref{sec:results}, homogeneous periodic sequences generically fail to produce a genuine Parrondo reversal.
\subsection{Phase-induced spatial inhomogeneity and scattering decomposition}
\label{sec:methods_defect}
To enable directed transport beyond what band-structure engineering alone can achieve, we break the translational symmetry by introducing a localized phase defect at the origin. The position-dependent coin operator is defined as
\begin{equation}
\hat{C}_{\mathrm{def}}(\phi) = \sum_{x \neq 0} \ket{x}\!\bra{x} \otimes \hat{C}_0 + \ket{0}\!\bra{0} \otimes e^{i\phi}\hat{C}_0,
\label{eq:C_defect}
\end{equation}
where $\phi \in [0,2\pi)$ is the defect phase and $\hat{C}_0$ denotes the bulk coin for the current game step ($\hat{C}_A$ or $\hat{C}_B$). The full step operator $\hat{U}(\phi) = \hat{S}\,\hat{C}_{\mathrm{def}}(\phi)$ can be decomposed into a translationally invariant bulk contribution and a localized scattering term:
\begin{equation}
\hat{U}(\phi) = \hat{U}_{\mathrm{homo}} + \hat{V}(\phi),
\label{eq:U_split}
\end{equation}
where $\hat{U}_{\mathrm{homo}} = \hat{S}(\sum_x \ket{x}\!\bra{x} \otimes \hat{C}_0)$ is the standard homogeneous walk, and the scattering perturbation is
\begin{equation}
\hat{V}(\phi) = \hat{S}\Big[\ket{0}\!\bra{0} \otimes (e^{i\phi}-1)\hat{C}_0\Big].
\label{eq:V_def}
\end{equation}
For $\phi = 0$, the perturbation vanishes ($\hat{V} = 0$) and the full translational symmetry is restored.
The symmetry-breaking mechanism is most transparent in momentum space. The homogeneous component is diagonal in $k$, i.e., $\bra{k'}\hat{U}_{\mathrm{homo}}\ket{k} \propto \delta(k-k')$, whereas the defect term couples different momenta. Using $\braket{x\!=\!0|k} = 1/\sqrt{2\pi}$ from Eq.~\eqref{eq:k_basis} and the definition of $\hat{V}(\phi)$ in Eq.~\eqref{eq:V_def}, the coin-space matrix element between momentum states $\ket{k}$ and $\ket{k'}$ evaluates to
\begin{equation}
\bra{k'}\hat{V}(\phi)\ket{k}\big|_{\mathcal{H}_C}
= \frac{1}{2\pi}\,\bm{\Sigma}(k')\,(e^{i\phi}-1)\,\hat{C}_0,
\label{eq:k_mix}
\end{equation}
where the notation $|_{\mathcal{H}_C}$ emphasizes that the result is a $2\times 2$ matrix acting on the coin space. Crucially, this expression is independent of $k-k'$: the Dirac delta $\delta(k-k')$ that would enforce momentum conservation in the homogeneous case is absent. The defect therefore scatters an incoming mode at momentum $k$ into a continuum of outgoing modes $k'$, providing the momentum mixing necessary for rectification.
\subsection{Scattering observables and resonance-type phase response}
\label{sec:methods_scatt_obs}
The momentum-mixing matrix element in Eq.~\eqref{eq:k_mix} establishes that the phase defect acts as a scattering center. To quantify how this scattering redistributes probability current between the left- and right-moving channels---and thereby generates the directed transport underlying the Parrondo effect---we now formulate the problem in the language of single-defect scattering theory.
Far from the defect ($|x| \gg 1$), the dynamics are governed by the bulk Floquet operator $\bm{\mathcal{U}}_{\mathrm{eff}}(k)$, whose eigenstates $\ket{u_k^\pm}$ define right-moving ($v_g^{(+)}(k) > 0$) and left-moving ($v_g^{(-)}(k) < 0$) Bloch modes. A wave packet incident on the defect from either side is partially transmitted and partially reflected. We encode this mode conversion in a $2\times 2$ scattering matrix that relates the incoming and outgoing amplitudes at a fixed quasi-energy:
\begin{equation}
\begin{pmatrix} \psi_{\mathrm{out}}^{-} \\ \psi_{\mathrm{out}}^{+} \end{pmatrix}
=
\underbrace{
\begin{pmatrix} r(k,\phi) & t'(k,\phi) \\ t(k,\phi) & r'(k,\phi) \end{pmatrix}
}_{\displaystyle \bm{\mathcal{S}}(k,\phi)}
\begin{pmatrix} \psi_{\mathrm{in}}^{+} \\ \psi_{\mathrm{in}}^{-} \end{pmatrix}.
\label{eq:S_matrix}
\end{equation}
Here, $\psi_{\mathrm{in}}^{+}$ ($\psi_{\mathrm{in}}^{-}$) denotes a right-moving (left-moving) wave incident on the defect from the left (right), while $\psi_{\mathrm{out}}^{+}$ ($\psi_{\mathrm{out}}^{-}$) is the transmitted (reflected) outgoing amplitude. The off-diagonal elements $t(k,\phi)$ and $t'(k,\phi)$ are the transmission amplitudes for left-to-right and right-to-left incidence, respectively, while $r(k,\phi)$ and $r'(k,\phi)$ are the corresponding reflection amplitudes. The directional transmission probabilities are
\begin{equation}
\mathcal{T}_{L\to R}(k,\phi) = |t(k,\phi)|^2, \qquad
\mathcal{T}_{R\to L}(k,\phi) = |t'(k,\phi)|^2.
\label{eq:T_LR_def}
\end{equation}
The key quantity for rectification is the \emph{transmission asymmetry}:
\begin{equation}
\Delta \mathcal{T}(k,\phi) \equiv \mathcal{T}_{L\to R}(k,\phi) - \mathcal{T}_{R\to L}(k,\phi).
\label{eq:DeltaT_def}
\end{equation}
A nonzero $\Delta\mathcal{T}$ means the defect transmits waves preferentially in one direction, which is the microscopic origin of the net current that drives the Parrondo reversal.
We now derive the functional form of $\mathcal{T}(k,\phi)$ using the transfer-matrix method. Consider a Bloch mode at quasi-momentum $k$ propagating through the bulk and encountering the defect at $x=0$. In the bulk, a single Floquet period maps the coin-space spinor via $\bm{\mathcal{U}}_{\mathrm{eff}}(k)$, while at the defect site the coin acquires the additional phase $e^{i\phi}$. The transfer matrix across the defect can be written as
\begin{equation}
\bm{M}(k,\phi) = \bm{\mathcal{U}}_{\mathrm{eff}}^{-1}(k)\;\bm{\Sigma}(k)\,e^{i\phi}\hat{C}_0\;,
\label{eq:transfer_matrix}
\end{equation}
which relates the spinor amplitudes on the two sides of $x=0$. Writing $\bm{M}$ in the Bloch eigenbasis $\{\ket{u_k^+},\ket{u_k^-}\}$, the transmission amplitude for left-to-right incidence is $t(k,\phi) = 1/M_{22}(k,\phi)$ (this is the standard transfer-matrix relation for single-defect scattering~\cite{PhysRevA.105.042216}). Since $\bm{M}$ depends on $\phi$ only through the overall factor $e^{i\phi}$, the modulus squared $|M_{22}|^2$ can be decomposed into a constant part and a term oscillating in $\phi$, yielding the generic form
\begin{equation}
|M_{22}(k,\phi)|^2 = 1 + \mathcal{F}(k)\sin^2\!\!\left(\frac{\phi - \phi_{\mathrm{res}}(k)}{2}\right),
\label{eq:M22_squared}
\end{equation}
yielding the Fabry--P\'{e}rot-like resonance formula for the transmission probability:
\begin{equation}
\mathcal{T}_{L\to R}(k,\phi) = |t(k,\phi)|^2 = \frac{1}{1+\mathcal{F}(k)\sin^2\!\!\left(\frac{\phi-\phi_{\mathrm{res}}(k)}{2}\right)}.
\label{eq:T_resonance}
\end{equation}
The two parameters in this expression have clear physical origins:
\begin{itemize}
\item \emph{Resonance phase} $\phi_{\mathrm{res}}(k)$: determined by the argument of the off-diagonal Bloch matrix elements of $\hat{C}_0$ at momentum $k$. At $\phi = \phi_{\mathrm{res}}(k)$, the defect-induced phase exactly compensates the bulk dynamical phase, and perfect transmission ($\mathcal{T}=1$) is achieved.
\item \emph{Finesse} $\mathcal{F}(k)$: controlled by the coin mixing strength $\beta$. Explicitly, $\mathcal{F}(k) \propto \sin^2(2\beta)\,f(k,\alpha,\gamma)$, where $f$ encodes the $k$-dependent interference geometry. In the non-mixing limits $\beta \to 0$ or $\beta \to \pi/2$, one of the coin-state channels decouples and $\mathcal{F}(k) \to 0$, rendering $\mathcal{T}$ insensitive to $\phi$ and thereby suppressing rectification.
\end{itemize}
An analogous expression holds for $\mathcal{T}_{R\to L}(k,\phi)$ with a generally different resonance phase, which is the origin of the directional asymmetry $\Delta\mathcal{T} \neq 0$.
\begin{figure}[h]
    \centering
    \includegraphics[width=\linewidth]{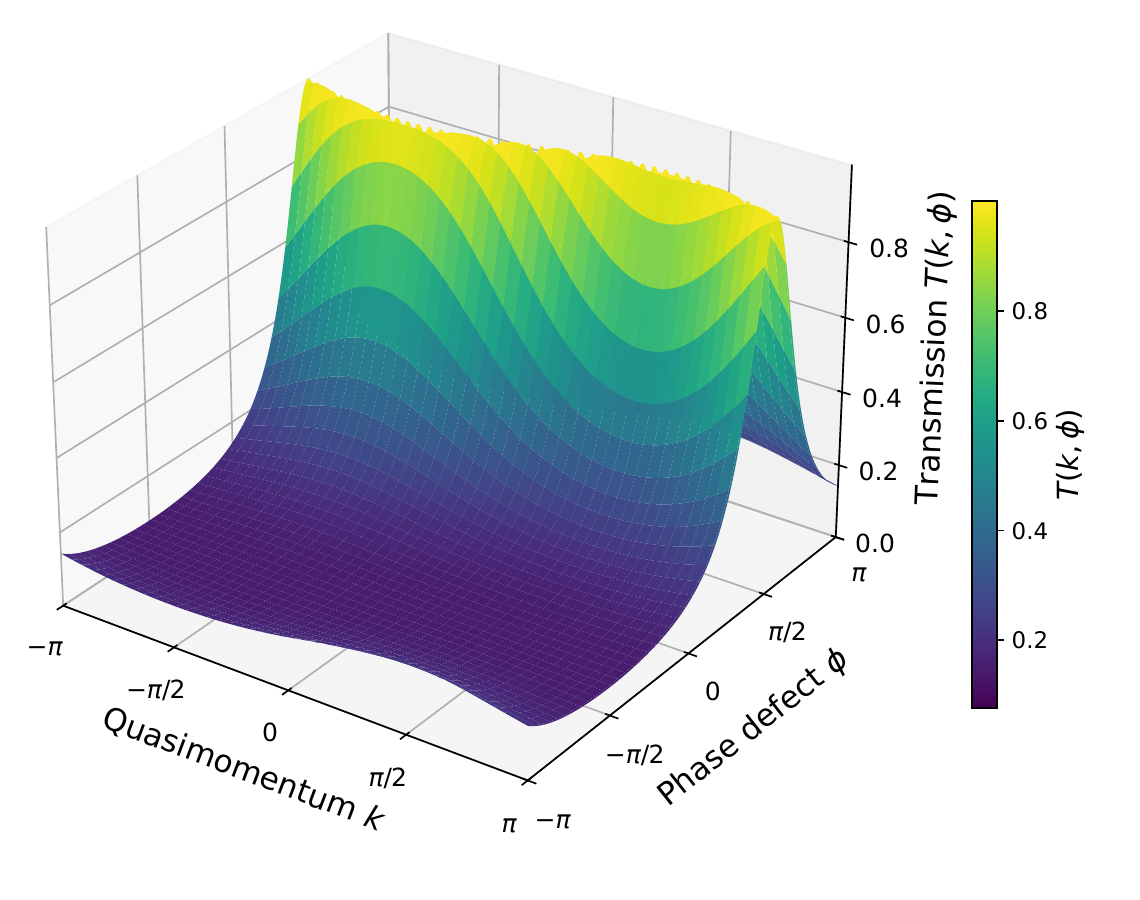}
    \caption{\textbf{Resonance-type phase response of the transmission probability.} The transmission $\mathcal{T}_{L\to R}(k,\phi)$ is plotted as a function of quasi-momentum $k$ and defect phase $\phi$, exhibiting the characteristic Fabry--P\'{e}rot-like resonance structure described by Eq.~\eqref{eq:T_resonance}. Perfect transmission ($\mathcal{T}=1$) occurs along the resonance curve $\phi = \phi_{\mathrm{res}}(k)$, while the width of the transmission dip is governed by the finesse $\mathcal{F}(k)$.}
    \label{fig:T_resonance_3D}
\end{figure}
\subsection{Fourier structure of the asymptotic drift}
\label{sec:methods_fourier}
The decomposition $\hat{U}(\phi) = \hat{U}_{\mathrm{homo}} + \hat{V}(\phi)$ [Eq.~\eqref{eq:U_split}] with $\hat{V}(\phi) \propto (e^{i\phi}-1)$ [Eq.~\eqref{eq:V_def}] implies a useful polynomial structure for the time-evolved state. Since $\hat{U}(\phi) = \hat{U}_{\mathrm{homo}} + (e^{i\phi}-1)\hat{W}$, where $\hat{W} \equiv \hat{S}[\ket{0}\!\bra{0}\otimes\hat{C}_0]$ is $\phi$-independent, the $t$-step propagator $\hat{U}(\phi)^t$ can be expanded by collecting all terms with exactly $m$ factors of $(e^{i\phi}-1)$:
\begin{equation}
\ket{\Psi(t;\phi)} = \hat{U}(\phi)^t\ket{\Psi(0)}
= \sum_{m=0}^{t} (e^{i\phi}-1)^m\,\ket{\Psi^{(m)}(t)},
\label{eq:psi_poly}
\end{equation}
where each coefficient state $\ket{\Psi^{(m)}(t)}$ is independent of $\phi$ and corresponds to all propagation histories with exactly $m$ scattering events at the defect among the $t$ time steps, with the remaining $t-m$ steps governed by the homogeneous unitary $\hat{U}_{\mathrm{homo}}$.
We quantify the transport through the position expectation value:
\begin{equation}
\langle \hat{x} \rangle(t;\phi) = \bra{\Psi(t;\phi)}\,\hat{x}\,\ket{\Psi(t;\phi)}.
\label{eq:x_expect}
\end{equation}
Since this quantity is quadratic in the state, it inherits a finite Fourier series in $\phi$ at each time step:
\begin{equation}
\langle \hat{x} \rangle(t;\phi) = \sum_{n=-t}^{t} c_n(t)\,e^{in\phi},
\label{eq:x_fourier_finite}
\end{equation}
with time-dependent Fourier coefficients $c_n(t)$. Defining the asymptotic drift velocity through the long-time average,
\begin{equation}
v(\phi) = \lim_{T\to\infty} \frac{1}{T} \sum_{t=0}^{T-1} \frac{\langle \hat{x} \rangle(t;\phi)}{t},
\label{eq:v_asymptotic}
\end{equation}
the $2\pi$-periodicity in $\phi$ yields the Fourier expansion
\begin{equation}
v(\phi) = \sum_{n=-\infty}^{\infty} \bar{c}_n\, e^{in\phi},
\qquad \bar{c}_n \equiv \lim_{T\to\infty} \frac{1}{T}\sum_{t=0}^{T-1} \frac{c_n(t)}{t}.
\label{eq:v_complex_fourier}
\end{equation}
Since $v(\phi) \in \mathbb{R}$, the coefficients satisfy $\bar{c}_{-n} = \bar{c}_n^*$, and the expansion can be recast as a real sine--cosine series:
\begin{equation}
v(\phi) = v_{\mathrm{bulk}} + \sum_{n=1}^{\infty} A_n \sin(n\phi - \delta_n),
\label{eq:v_sin_series}
\end{equation}
where $v_{\mathrm{bulk}} \equiv v(\phi\!=\!0)$ is the drift in the absence of the defect, $A_n \geq 0$ are the harmonic amplitudes, and $\delta_n$ are the corresponding phase offsets. Equation~\eqref{eq:v_sin_series} is the most general $\phi$-dependence consistent with a single local phase defect; in practice, the series is well-approximated by a finite truncation when the higher harmonics decay sufficiently fast.
To formalize the Parrondo condition, we define the \emph{rectification ratio}
\begin{equation}
\mathcal{R}(\phi)
\equiv
\frac{\sum_{n=1}^{\infty} A_n \sin(n\phi-\delta_n)}{|v_{\mathrm{bulk}}|},
\label{eq:Rphi}
\end{equation}
which measures the strength of the defect-induced drift relative to the intrinsic bulk bias. From Eq.~\eqref{eq:v_sin_series}, the total drift can be written as
\begin{equation}
v(\phi) = v_{\mathrm{bulk}} + |v_{\mathrm{bulk}}|\,\mathcal{R}(\phi)\,\mathrm{sgn}(v_{\mathrm{bulk}}).
\end{equation}
For a losing game ($v_{\mathrm{bulk}} < 0$), this gives $v(\phi) = v_{\mathrm{bulk}}[1 - \mathcal{R}(\phi)]$, so the combined drift becomes positive when $\mathcal{R}(\phi) > 1$---that is, when the defect-induced rectification overcomes the bulk losing tendency. We thus identify $\mathcal{R}(\phi) > 1$ as the \emph{winning} (Parrondo) regime.
\subsection{Bulk-plus-rectification decomposition}
\label{sec:methods_master}
We now connect the drift velocity $v(\phi)$ directly to the scattering asymmetry $\Delta\mathcal{T}$ derived above. The physical reasoning is as follows: in the asymptotic ($t\to\infty$) regime, the walker's drift arises from two contributions: (i)~the group-velocity transport in each Bloch band, which persists even without the defect, and (ii)~the redistribution of probability current between bands caused by scattering at the defect. The second contribution is proportional to the transmission asymmetry $\Delta\mathcal{T}(k,\phi)$, weighted by how strongly each mode $k$ is populated. To formalize this, we define the momentum-resolved net transmission weighted by the initial state:
\begin{equation}
\mathcal{T}_{\mathrm{net}}(\phi) \equiv
\int_{-\pi}^{\pi}\!dk\;\Delta \mathcal{T}(k,\phi)\,\mathcal{W}(k),
\label{eq:Tphi_def}
\end{equation}
where $\mathcal{W}(k) \equiv \sum_\sigma |\!\braket{u_k^\sigma|\chi_c}\!|^2/(2\pi)$ is the spectral weight of the initial coin state $\ket{\chi_c}$ projected onto the Bloch eigenmodes. The drift velocity then admits a transparent bulk-plus-rectification form:
\begin{equation}
v(\phi) = v_{\mathrm{bulk}}
+ \int_{-\pi}^{\pi}\!dk\;
\bar{v}_g(k)\,\Delta \mathcal{T}(k,\phi)\,\mathcal{W}(k),
\label{eq:v_bulk_plus_rect}
\end{equation}
where $\bar{v}_g(k) \equiv [v_g^{(+)}(k) - v_g^{(-)}(k)]/2$ is the mean group-velocity magnitude at momentum $k$. The first term captures the drift inherent to the bulk coin sequence (which is negative for a losing game), while the integral represents the rectification current generated by the defect-induced scattering asymmetry $\Delta\mathcal{T}$. A positive Parrondo effect requires the rectification integral to exceed $|v_{\mathrm{bulk}}|$ in magnitude.
\section{Results}
\label{sec:results}
\subsection{Demonstration of the paradox with general coins}
\label{sec:numerical_results}
\begin{figure*}[t]
    \centering
    \subfloat[\textbf{General coin parameters.}
    The panels display the space-time evolution of the probability distribution (blue density) and $\langle \hat{x} \rangle$ (solid black line).
    \textbf{Top Row:} Homogeneous lattice ($\phi=0$). \textbf{Bottom Row:} Inhomogeneous lattice ($\phi=\pi/2$) at the origin (dashed yellow line).
    For the parameters in Eq.~\eqref{eq:selected_params}, Games~A and~B show negative drift, while Sequence ABB reverses to positive drift.\label{fig:parrondo_general}]{%
        \includegraphics[width=0.85\linewidth]{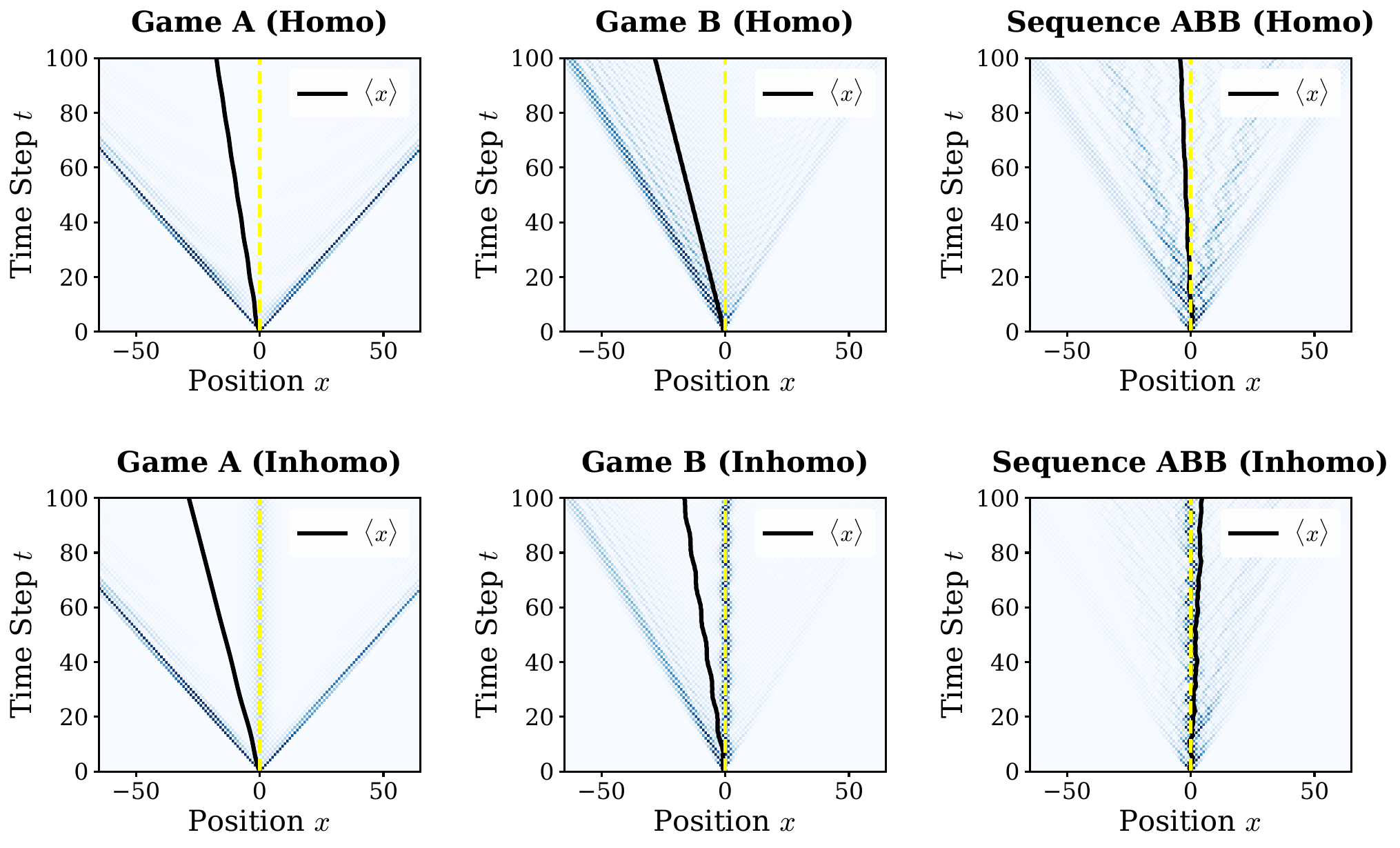}}\\[6pt]
    \subfloat[\textbf{Standard coin angles} ($\alpha, \gamma \in \{0, \pi/2, \pi\}$).
    Parameters: $\alpha=0$, $\gamma=\pi/2$, $\beta_C \approx 6.12$, $\beta_D \approx 2.26$.
    The paradox persists: Games~C and~D exhibit negative drift, while Sequence CDD yields positive drift.\label{fig:parrondo_standard}]{%
        \includegraphics[width=0.85\linewidth]{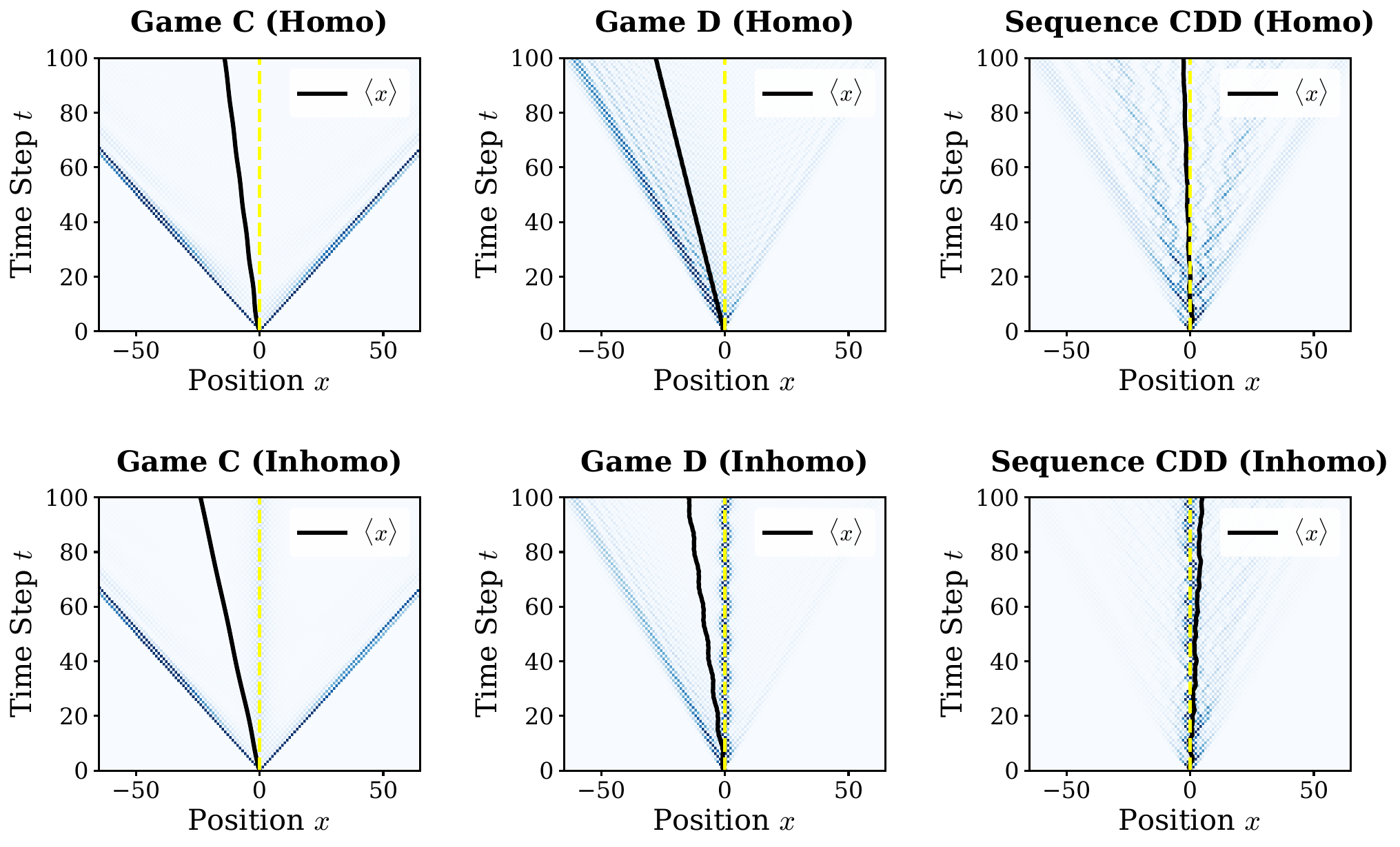}}
    \caption{\textbf{Demonstration of Quantum Parrondo's Paradox.} Space-time evolution of the probability distribution (blue density) and mean position $\langle \hat{x} \rangle$ (solid black line) for (a)~general coin parameters (Games~A and~B, Sequence~ABB) and (b)~standard coin parameters (Games~C and~D, Sequence~CDD). In both cases, the individual games lose ($\langle \hat{x} \rangle < 0$) while the combined sequence wins ($\langle \hat{x} \rangle > 0$) only in the presence of the phase defect.}
\end{figure*}

To demonstrate the emergence of Parrondo's paradox in the quantum walk dynamics, we employ the general $SU(2)$ coin operator~\eqref{eq:SU2}. The specific coin parameters used in our simulations were identified through a systematic parameter scan over $(\alpha,\gamma,\beta_A,\beta_B)$, as detailed in Appendix~\ref{app:parameter_scan}. The search space for the phase angles $\alpha$ and $\gamma$ was discretized in increments of $\pi/8$ to locate regions where the paradoxical inversion is most pronounced. Based on this analysis, we selected the following parameter set:
\begin{equation}
    \alpha = \frac{3\pi}{8}, \quad \gamma = \frac{\pi}{8}, \quad \beta_A \approx 6.07, \quad \beta_B \approx 5.42.
    \label{eq:selected_params}
\end{equation}
The resulting space-time evolution of the probability distribution is shown in Fig.~\ref{fig:parrondo_general}. Under these conditions, the individual games A and B exhibit a negative drift (loss) in the presence of a phase defect, as indicated by the leftward trajectory of $\langle \hat{x} \rangle$. However, the alternating sequence ABB produces a positive drift (win), clearly demonstrating the paradoxical behavior. The quantitative values of $\langle \hat{x} \rangle$ at $N=100$ steps are summarized in Table~\ref{tab:drift_summary}.

\begin{figure*}[t]
    \centering
    \includegraphics[width=0.85\textwidth]{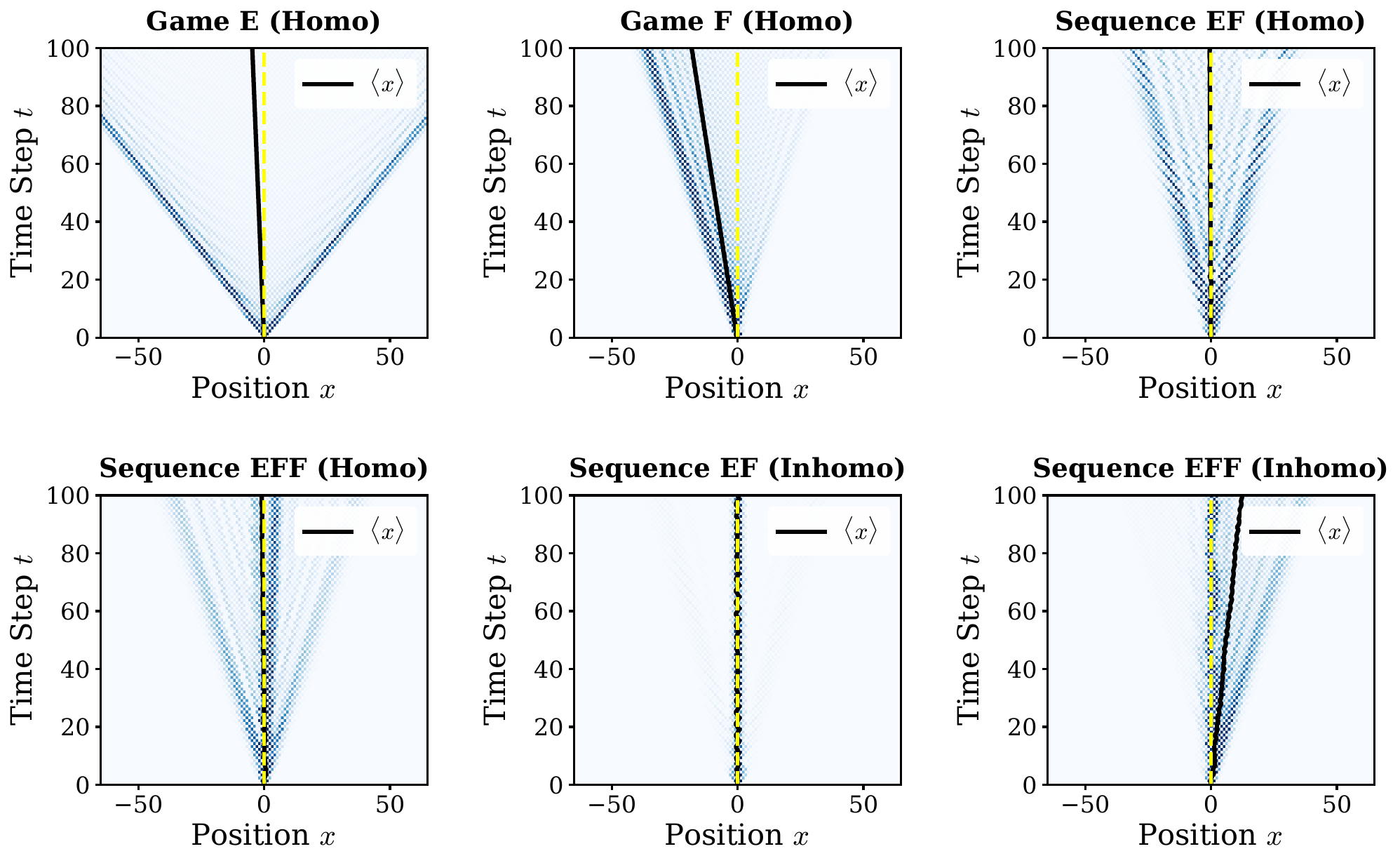}
    \caption{\textbf{Space-time probability distributions of quantum walks under different game sequences (Games~E and~F).} The blue gradient indicates probability density, and the solid black line represents the expectation value $\langle \hat{x} \rangle$. \textbf{Top row:} Homogeneous sequences ($\phi=0$). \textbf{Bottom row:} Sequences with an inhomogeneous phase defect ($\phi=\pi/2$) at the origin.}
    \label{fig:qw_spacetime}
\end{figure*}

To verify the robustness of the mechanism and connect our findings with established quantum walk models (which often utilize discrete angles like $0$ or $\pi/2$), we further investigated the standard parameter subspace where $\alpha, \gamma \in \{0, \pi/2, \pi\}$. The global drift velocity landscapes for these standard cases are detailed in Appendix~\ref{app:parameter_scan} (see Figs.~\ref{fig:scan_A_std}--\ref{fig:scan_ABB_std}). Although the standard angle combinations typically yield narrower regions of paradoxical behavior compared to the fully optimized general angles, a prominent paradoxical configuration was identified at:
\begin{equation}
    \alpha = 0, \quad \gamma = \frac{\pi}{2}, \quad \beta_C \approx 6.12, \quad \beta_D \approx 2.26.
    \label{eq:standard_params}
\end{equation}
The space-time evolution for this configuration is presented in Fig.~\ref{fig:parrondo_standard}. Consistent with the general case, the individual games C and D exhibit a negative drift ($\langle \hat{x} \rangle < 0$) in the presence of the phase defect ($\phi=\pi/2$), while the alternating sequence CDD successfully reverses this trend, resulting in a positive drift ($\langle \hat{x} \rangle > 0$). This confirms that the phase-defect-induced Parrondo's paradox is a generic feature of these quantum walks; the symmetry-breaking mechanism remains effective even when the coin phases are restricted to standard discrete values.

\subsection{Reassessment of the winning criterion}
\label{sec:winning_criterion}
We next reassess the winning criterion by comparing with the coin parameters adopted from the experimental realization by Jan \textit{et al.}~\cite{jan2020experimental}. The coin operators are parameterized in the general $SU(2)$ form~\eqref{eq:SU2}, with parameters: $\alpha_E = 2.395$, $\beta_E = 0.513$, $\gamma_E = 0.909$ for Game~E, and $\alpha_F = 2.611$, $\beta_F = 1.176$, $\gamma_F = 2.313$ for Game~F.
In the reference study~\cite{jan2020experimental}, the occurrence of Parrondo's paradox was characterized by the probability asymmetry $P_R - P_L > 0$, defined as the difference between the probability of finding the walker in the right and left half-spaces. However, our theoretical analysis reveals that a positive probability asymmetry does not strictly correspond to a ``winning'' outcome in terms of capital growth. Due to the broad dispersion of the quantum wave packet, it is possible for a state to satisfy $P_R - P_L > 0$ while simultaneously having a negative expectation value $\langle \hat{x} \rangle < 0$. We therefore adopt the position expectation value $\langle \hat{x} \rangle$ as the rigorous criterion for evaluating the game's outcome, consistent with the classical definition in Eq.~\eqref{eq:losing_games}.
The space-time evolution of the probability density under these parameters is illustrated in Fig.~\ref{fig:qw_spacetime}. The top row displays the results for homogeneous quantum walks: both individual games (E and F) and their periodic sequences (EF and EFF) exhibit a negative drift, consistent with the bias of the individual coins. The bottom row demonstrates the emergence of the paradox under inhomogeneous conditions---the phase defect reverses the direction of the drift for the combined sequences EF and EFF, while the individual games remain losing. This result highlights that the ``winning'' behavior is not solely a property of the coin sequence but arises from the interplay between the sequence order and the spatial inhomogeneity of the system. The complete numerical results for all three parameter sets are compiled in Table~\ref{tab:drift_summary}.

\begin{table}[!htbp]
    \caption{\textbf{Asymptotic expected position $\langle \hat{x} \rangle$ at $N=100$ steps.} Results are shown for all coin parameter sets under homogeneous ($\phi=0$) and inhomogeneous ($\phi=\pi/2$) conditions. Bold positive values indicate the paradoxical winning regime where the combined sequence reverses the drift direction.}
    \label{tab:drift_summary}
    \begin{ruledtabular}
    \begin{tabular}{llrr}
    \textrm{Parameter set} & \textrm{Sequence} & \textrm{$\phi=0$} & \textrm{$\phi=\pi/2$} \\
    \colrule
    General (A/B) & Game A   & $-17.50$ & $-28.63$ \\
                  & Game B   & $-28.40$ & $-16.58$ \\
                  & Seq.\ ABB & $-4.32$  & $\mathbf{+4.59}$ \\
    \colrule
    Standard (C/D) & Game C   & $-14.32$ & $-24.08$ \\
                   & Game D   & $-28.04$ & $-15.02$ \\
                   & Seq.\ CDD & $-2.75$  & $\mathbf{+4.99}$ \\
    \colrule
    Jan \textit{et al.}\ (E/F) & Game E   & $-4.72$  & $-5.85$ \\
                               & Game F   & $-18.24$ & $-1.31$ \\
                               & Seq.\ EF  & $-0.49$  & $-0.09$ \\
                               & Seq.\ EFF & $-1.03$  & $\mathbf{+12.56}$ \\
    \end{tabular}
    \end{ruledtabular}
\end{table}

\subsection{Phase-controlled transport and winning strategies}
\label{sec:phase_controlled}
\begin{figure}[!htbp]
    \centering
    \includegraphics[width=\linewidth]{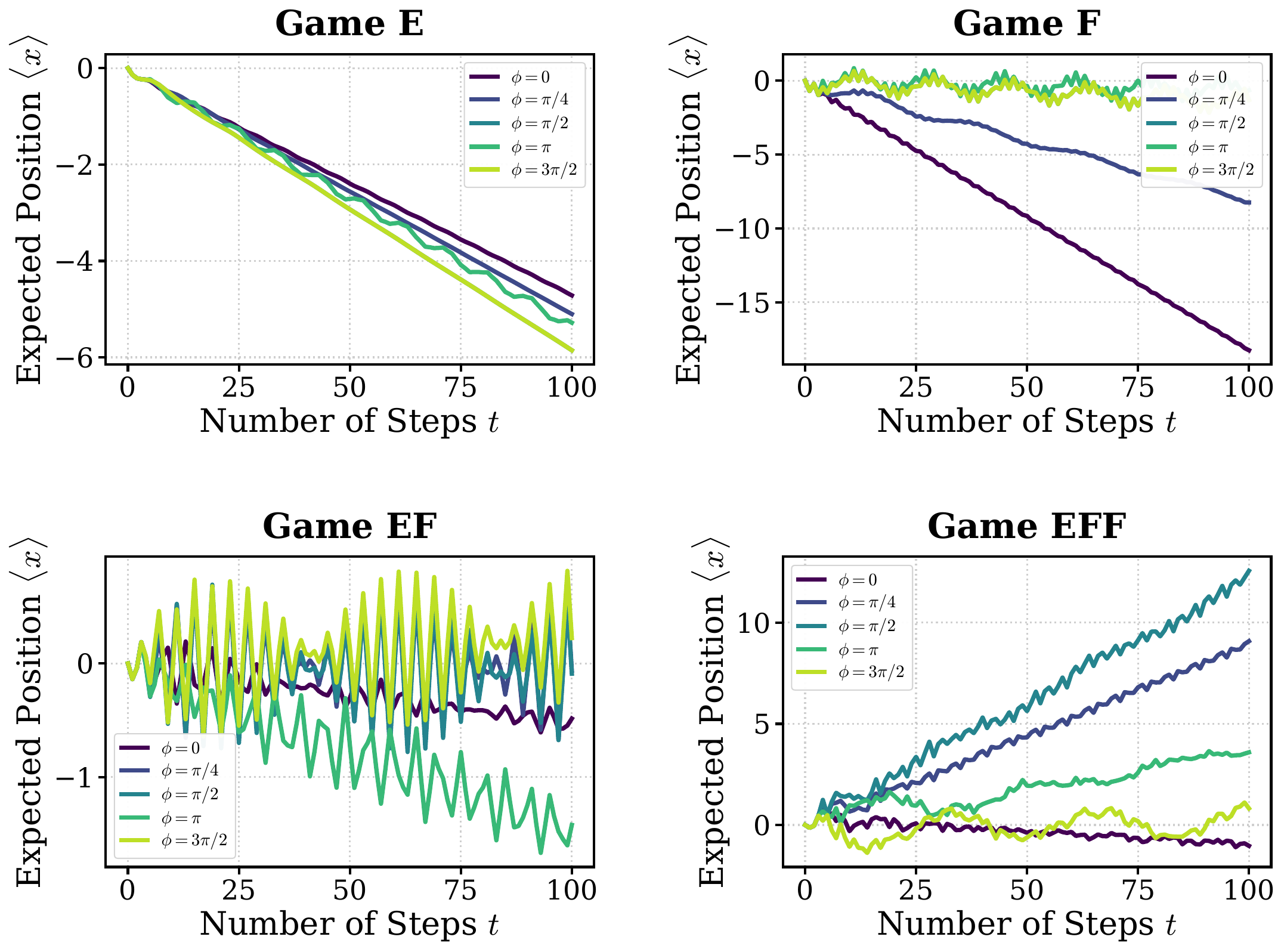}
    \caption{Evolution of the expected position $\langle \hat{x} \rangle$ as a function of time steps $t$ for different game sequences: (a) Game~E, (b) Game~F, (c) Sequence EF, and (d) Sequence EFF. The various curves correspond to different phase defects $\phi$ ranging from $0$ to $3\pi/2$. Homogeneous cases ($\phi=0$, dark purple) consistently show negative drift. For sequences EF and EFF, a phase defect of $\phi=\pi/2$ (teal) or $\phi=3\pi/2$ (yellow-green) reverses the drift to positive values, demonstrating the phase-controlled quantum Parrondo's paradox.}
    \label{fig:game_phase_comparison_large}
\end{figure}

To systematically investigate the impact of the phase defect, we analyze the time evolution of $\langle \hat{x} \rangle_t$ [cf.\ Eq.~\eqref{eq:drift_def}] for different game sequences under varying phase parameters $\phi \in [0, 2\pi)$. Figure~\ref{fig:game_phase_comparison_large} illustrates the trajectories over $N=100$ steps for Game~E, Game~F, and the periodic sequences EF and EFF.
In the homogeneous limit ($\phi=0$), represented by the dark purple curves, all four sequences exhibit a negative drift, confirming the losing nature of the default coin parameters. For the individual games, the phase defect modifies the drift magnitude but does not qualitatively reverse the losing trend. However, for the combined sequences EF and EFF, the influence of the phase defect is dramatic: around $\phi \approx \pi/2$ and $\phi \approx 3\pi/2$, the expectation value becomes positive, with the walker exhibiting ballistic transport towards the right. This sensitivity to the local phase defect demonstrates that the Parrondo effect can be actively controlled by manipulating the phase at the origin.
In the classical regime, inducing a positive drift from two losing games typically relies on a noise-driven ratchet mechanism that competes against diffusive broadening. In the quantum regime, by contrast, the rectification term $\Delta\mathcal{T}(k, \phi)$ [Eq.~\eqref{eq:DeltaT_def}] originates from constructive interference, generating a robust and parameter-controllable ballistic current $v(\phi)$. Since $\langle \hat{x} \rangle(t) \approx v(\phi)\,t$, the quantum ratchet transports the walker over a distance $L \propto t$, compared with the classical diffusive scaling $L \propto \sqrt{t}$~\cite{aharonov1993quantum, kempe2003quantum}. The quantum Parrondo protocol thus exhibits both a quadratic speed-up in transport efficiency and a directed drift enabled by unitary scattering at the phase defect.
\subsection{Harmonic analysis of phase-controlled drift}
\label{sec:harmonic_analysis}
To quantify the long-term transport behavior, we extract the asymptotic drift velocity $v(\phi)$ [Eq.~\eqref{eq:drift_def}] from numerical data and decompose it into the Fourier series derived in Sec.~\ref{sec:methods_fourier} [Eq.~\eqref{eq:v_sin_series}]. In practice, the series is truncated at a finite order $N_{\mathrm{fit}}$, and the harmonic amplitudes $A_n$ and phase offsets $\delta_n$ are fit to the simulation data.

\begin{figure}[!htbp]
    \centering
    \includegraphics[width=\linewidth]{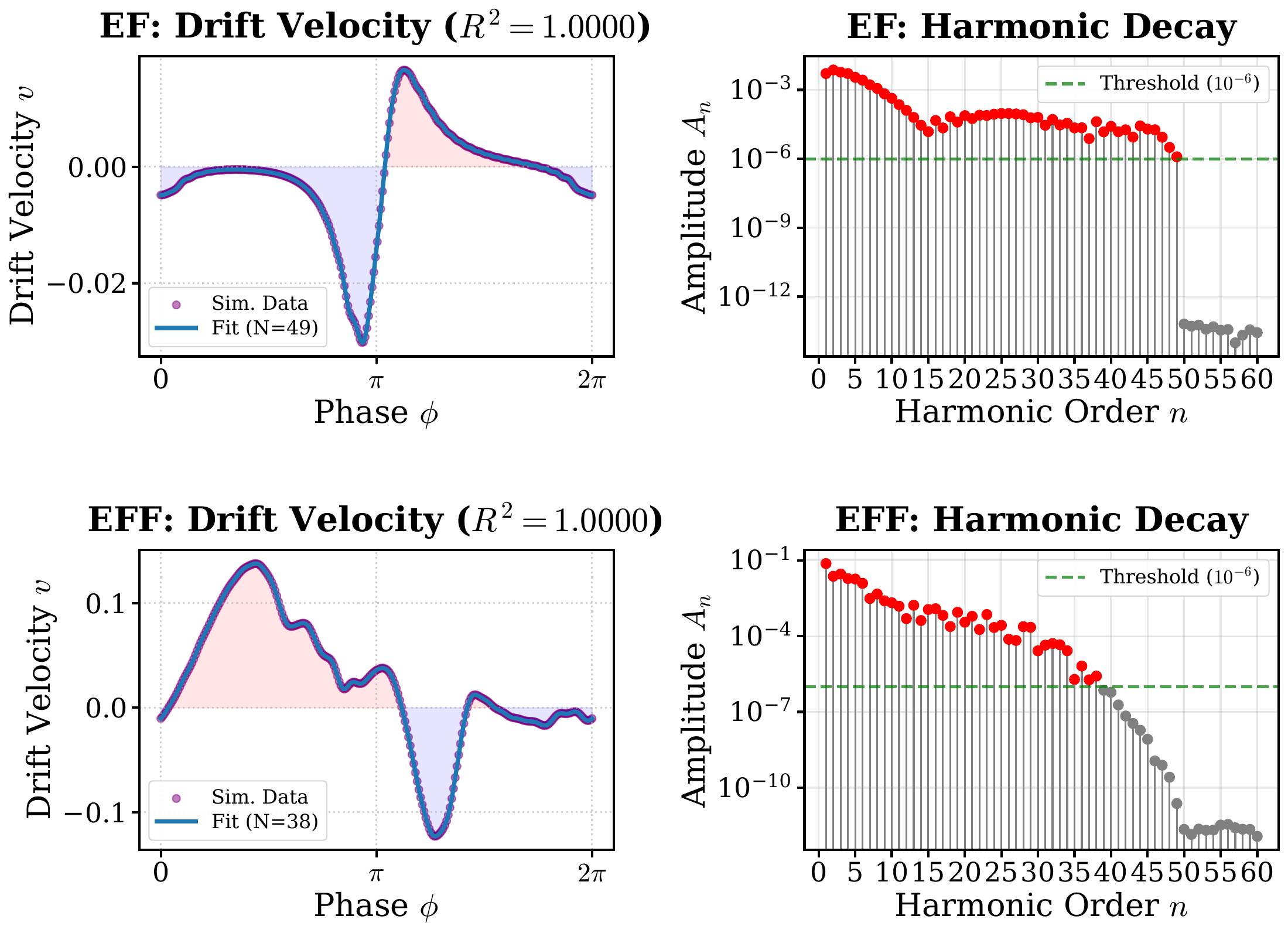}
    \caption{Drift velocity analysis and harmonic decomposition for sequences EF (top row) and EFF (bottom row). \textbf{Left column:} Drift velocity $v$ versus phase defect $\phi$. Purple dots: simulation data ($N=100$ steps); solid blue lines: Fourier fit. Red and blue shaded regions indicate winning ($v>0$) and losing ($v<0$) regimes. \textbf{Right column:} Amplitude spectrum $A_n$ on a logarithmic scale. The green dashed line marks the truncation threshold ($10^{-6}$). Significant amplitudes persist up to $n \approx 50$, reflecting the complex interference structure.}
    \label{fig:harmonic_analysis}
\end{figure}

Figure~\ref{fig:harmonic_analysis} presents the harmonic analysis for sequences EF (top row) and EFF (bottom row). The shaded red regions highlight the phase windows where $v(\phi) > 0$, confirming the realization of the quantum Parrondo's paradox. For both sequences, a large number of harmonic terms (up to $n \approx 49$ for EF and $n \approx 38$ for EFF) are required to capture the fine features of the drift curve. This slow decay of high-order harmonics suggests that the localized phase defect induces highly nontrivial multi-path interference, which allows for sharp transitions between winning and losing regimes.

\subsection{Entanglement dynamics in winning and losing regimes}
\label{sec:entanglement}
To quantify the quantum correlations developed during the walk, we calculate the time-dependent von Neumann entropy between the coin and position degrees of freedom. Recalling the Hilbert-space convention $\mathcal{H} = \mathcal{H}_P \otimes \mathcal{H}_C$ [Eq.~\eqref{eq:hilbert}], the state of the walker at time $t$ can be expressed as
\begin{equation}
    |\Psi(t)\rangle = \sum_{x} |x\rangle_P \otimes \bigl( \psi_{0,x}(t)|0\rangle_C + \psi_{1,x}(t)|1\rangle_C \bigr),
\end{equation}
where $|0\rangle_C$ and $|1\rangle_C$ denote the coin basis states and $|x\rangle_P$ denotes the position states. The reduced density matrix of the coin subsystem is obtained by tracing over the position space:
\begin{equation}
    \rho_C(t) = \mathrm{Tr}_P \bigl( |\Psi(t)\rangle \langle \Psi(t)| \bigr) =
    \begin{pmatrix}
    \sum_x |\psi_{0,x}|^2 & \sum_x \psi_{0,x}\psi_{1,x}^* \\
    \sum_x \psi_{1,x}\psi_{0,x}^* & \sum_x |\psi_{1,x}|^2
    \end{pmatrix},
    \label{eq:reduced_rho}
\end{equation}

where the time argument $(t)$ on the amplitudes is suppressed for compactness. The entanglement entropy is then
\begin{equation}
    S_E(t) = - \mathrm{Tr}\bigl(\rho_C(t) \log_2 \rho_C(t)\bigr) = - \sum_{i=1}^{2} \lambda_i \log_2 \lambda_i,
    \label{eq:entropy_def}
\end{equation}

where $\lambda_i$ are the eigenvalues of $\rho_C(t)$. A value of $S_E = 0$ indicates a separable product state, while $S_E = 1$ corresponds to a maximally entangled state.
\begin{figure}[!htbp]
    \centering
    \includegraphics[width=\linewidth]{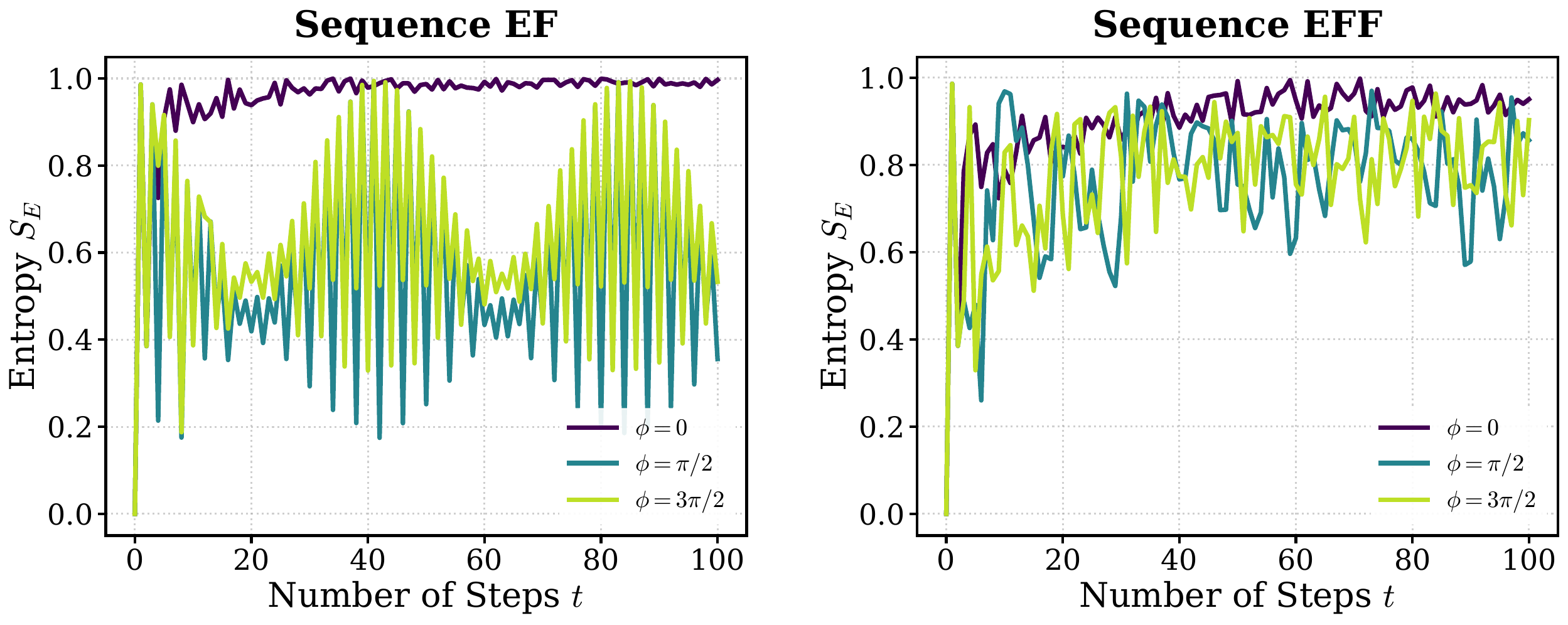}
    \caption{Time evolution of the von Neumann entropy $S_E(t)$ for sequences EF (left) and EFF (right) under different phase defects. The dark purple line ($\phi=0$) represents the homogeneous losing case, characterized by high and stable entanglement. The teal and yellow-green lines ($\phi=\pi/2, 3\pi/2$) correspond to the winning regimes. Notably, Sequence EF exhibits significant entropy oscillations in the winning phase, suggesting a cyclic restoration of coherence that supports ballistic transport.}
    \label{fig:entropy_evolution}
\end{figure}
\subsection{Initial state independence}
\label{sec:initial_state}
\begin{figure}[!htbp]
    \centering
    \includegraphics[width=\linewidth]{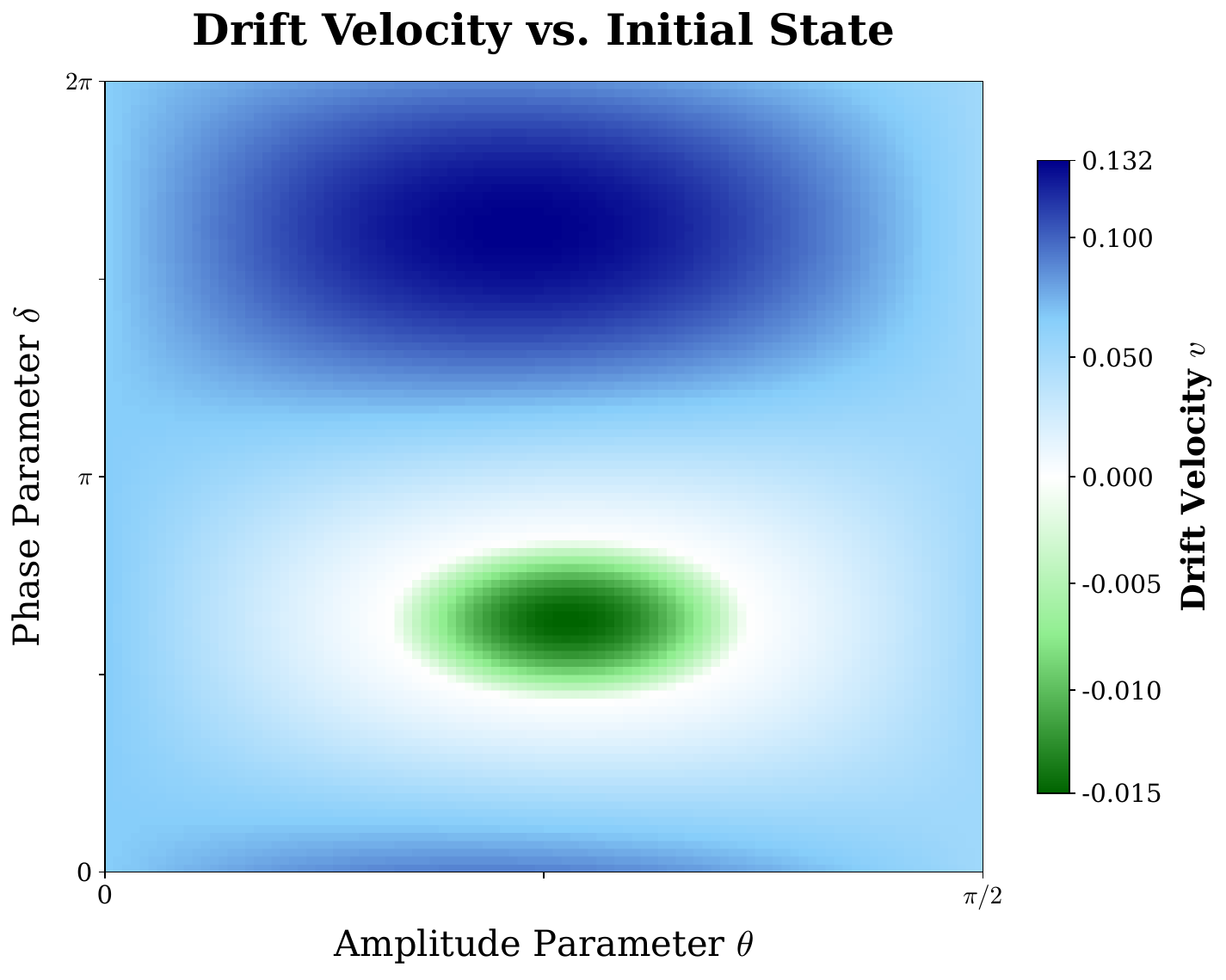}
    \caption{\textbf{Robustness of Drift Velocity against Initial State Parameters.}
    2D heatmap of the drift velocity $v$ as a function of the initial amplitude parameter $\theta$ and relative phase $\delta$.
    \textbf{Blue} shades represent positive drift (winning), \textbf{green} shades represent negative drift (losing), and \textbf{white} marks zero.
    The overwhelming prevalence of blue regions confirms that the phase-induced rectification mechanism is robust and largely independent of the specific initial coin state preparation.}
    \label{fig:initial_state_2d}
\end{figure}

To confirm the robustness of the ratchet effect against initial state variations, we performed a high-resolution parameter scan over the general initial state $\ket{\Psi_c(0)} = \cos\theta\ket{0} + e^{i\delta}\sin\theta\ket{1}$ using the Sequence~EFF coin parameters (Games~E and~F) with a phase defect $\phi=\pi/2$. Figure~\ref{fig:initial_state_2d} presents the asymptotic drift velocity as a 2D heatmap in the $(\theta, \delta)$ plane. The dominance of the blue (positive-drift) region demonstrates that the winning strategy is resilient to changes in the initial superposition; the drift velocity remains positive for the vast majority of $(\theta, \delta)$ combinations, with minor negative-drift regions confined to narrow phase windows.
\section{Conclusion}
\label{sec:conclusion}
In this work, we have demonstrated a minimal and robust realization of the quantum Parrondo's paradox within the framework of discrete-time quantum walks. By introducing a single localized phase defect $e^{i\phi}$ at the origin, we broke the discrete translational symmetry of the lattice, transforming the system from a momentum-conserving ballistic walker into a directed ``quantum ratchet.'' Our scattering decomposition [Eqs.~\eqref{eq:U_split}--\eqref{eq:k_mix}] reveals that this spatial inhomogeneity acts as a momentum-mixing center whose transmission asymmetry $\Delta\mathcal{T}(k,\phi)$ provides the microscopic rectification current underlying the Parrondo reversal [Eq.~\eqref{eq:v_bulk_plus_rect}].

A central contribution of our study is the critical refinement of the winning criterion for quantum Parrondo games. We have shown that the probability asymmetry $P_R - P_L$, widely employed in the literature~\cite{chandrashekar2011parrondo,jan2020experimental}, is an insufficient metric: a state can satisfy $P_R - P_L > 0$ while simultaneously having a negative expectation value $\langle \hat{x} \rangle < 0$. By adopting $\langle \hat{x} \rangle$ as the rigorous observable---consistent with the classical capital-based definition [Eq.~\eqref{eq:losing_games}]---we identified genuine paradoxical regimes where two individually losing games combine to produce a positive drift. This approach challenges previous single-qubit implementations~\cite{chandrashekar2011parrondo,jan2020experimental} and establishes that a persistent Parrondo effect requires spatial symmetry breaking rather than merely tuned $SU(2)$ sequences, in agreement with the recent inhomogeneous-coin framework of \citet{PhysRevA.110.052440} and the space-inhomogeneous analysis by \citet{walczak2025parrondo}.

Through harmonic analysis we showed that the drift velocity $v(\phi)$ requires up to $\sim\!50$ Fourier harmonics to converge (Fig.~\ref{fig:harmonic_analysis}), reflecting the highly nontrivial multi-path interference generated by a single point scatterer. The entanglement dynamics further reveal that winning strategies are associated with cyclic coherence restoration between the coin and position degrees of freedom, complementing the entanglement-territory analysis of \citet{jan2023territories} and the entanglement generation studies of \citet{panda2022entangled}. The robustness of the ratchet effect across a wide range of initial states (Fig.~\ref{fig:initial_state_2d}) and its successful mapping onto standard discrete coin angles confirm that the mechanism is a generic feature of spatially inhomogeneous quantum walks rather than a fine-tuned artifact.

Several directions for future investigation emerge from this work. First, the single-defect model naturally generalizes to \emph{multiple phase defects} distributed along the lattice, which could enable richer interference patterns and tunable multi-band rectification, analogous to the multi-defect Parrondo strategies recently explored in continuous-time quantum walks~\cite{ximenes2025ctqw_defect}. Second, extending the framework to \emph{two-dimensional lattices} and higher-dimensional tori~\cite{naves2023quantum2d,hosaka2024parrondo} would clarify whether the phase-defect mechanism can generate directed transport along controlled directions in multi-dimensional settings. Third, incorporating \emph{decoherence and open-system effects}~\cite{walczak2023noise,ding2012quantum} would establish the extent to which the ratchet survives under realistic noise conditions---a prerequisite for any experimental implementation. On the experimental side, the minimal requirements of our model---a single-qubit coin and a localized phase control at one lattice site---make it well suited for near-term quantum platforms in the NISQ era~\cite{preskill2018quantum}. Superconducting qubit arrays~\cite{yan2019strongly,gong2021quantum} offer site-resolved phase control through local flux biasing, while integrated photonic circuits~\cite{tang2018experimental} can implement position-dependent phase shifts via thermo-optic tuning. Trapped-ion~\cite{zahringer2010realization} and neutral-atom~\cite{karski2009quantum} platforms similarly provide the single-site addressability needed to engineer the phase defect. A successful experimental demonstration would not only validate the quantum Parrondo ratchet but also establish a resource-efficient building block for directed quantum transport and coherence-assisted energy harvesting in quantum networks.
\bibliography{references}
\clearpage
\onecolumngrid
\appendix
\section{Parameter Space Scanning}
\label{app:parameter_scan}
To identify robust regions exhibiting Parrondo's paradox, we performed a systematic numerical scan of the coin parameters under a fixed phase defect $\phi = \pi/2$. We calculated the asymptotic drift velocity $v$ over the full rotation range $\beta_1, \beta_2 \in [0, 2\pi]$ for discretized phase angles $\alpha$ and $\gamma$, where $\beta_1$ and $\beta_2$ denote the rotation angles of the two constituent coins. In the resulting heatmaps, blue regions indicate negative drift (loss), while red regions indicate positive drift (win). The paradox is identified where both individual games yield negative drift, yet the combined sequence yields positive drift. The general scan (Sec.~\ref{sec:general_scan}) identifies Games~A and~B, while the standard scan (Sec.~\ref{sec:standard_scan}) identifies Games~C and~D.
\subsection{General Parameter Scan ($\pi/8$ Increments)}
\label{sec:general_scan}
\begin{figure}[!htbp]
    \centering
    \includegraphics[width=0.58\linewidth]{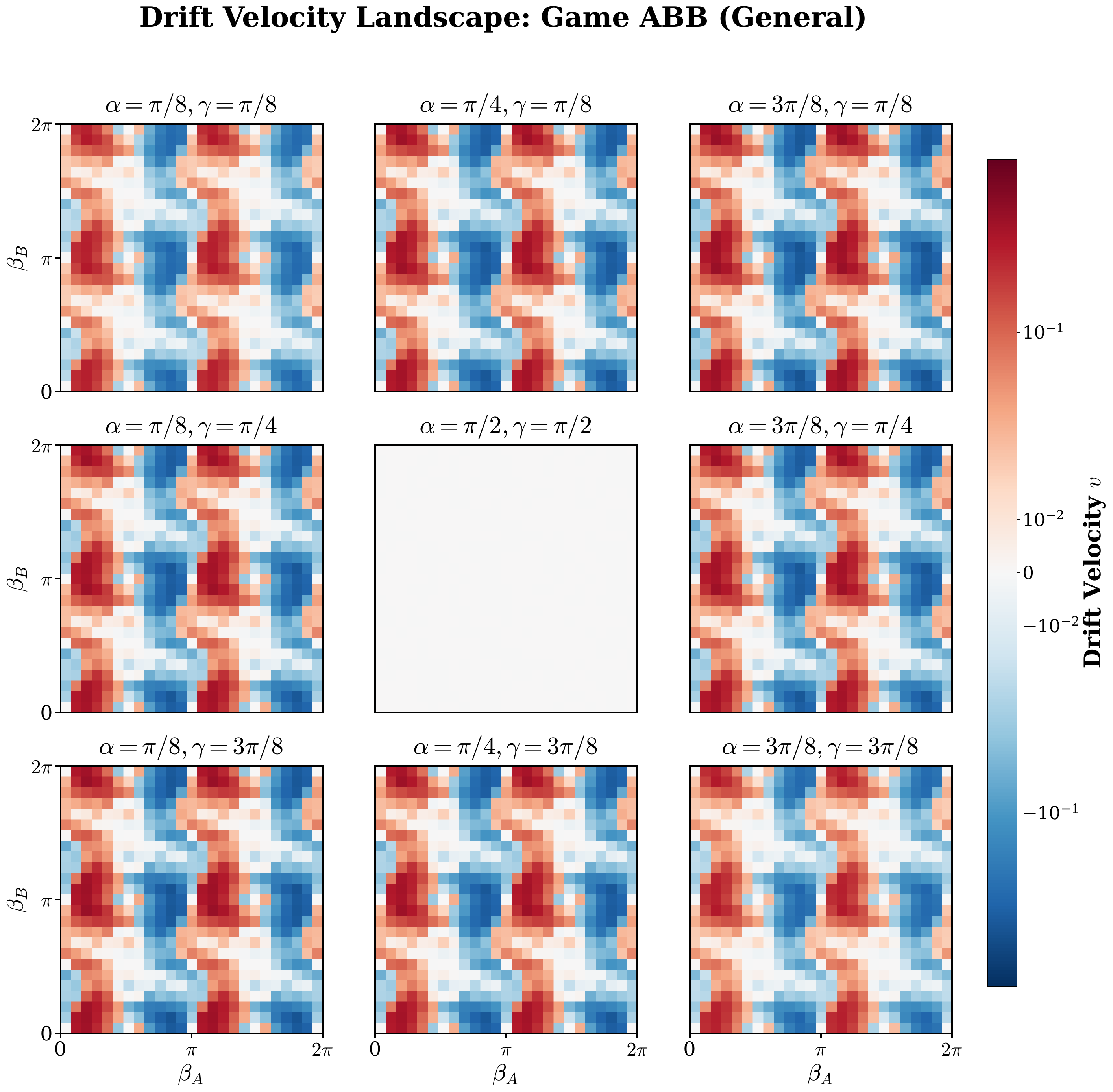}
    \caption{\textbf{Drift velocity landscape for Sequence ABB (General).} The optimal region ($\alpha=3\pi/8, \gamma=\pi/8$) shows a strong positive drift (red) emerging from the negative drifts of individual games.}
    \label{fig:scan_ABB}
\end{figure}

We first scanned $\alpha$ and $\gamma$ in increments of $\pi/8$. Based on the high-contrast regions observed in Fig.~\ref{fig:scan_ABB}, we selected the following optimized parameter set used in the main text:
\begin{equation}
    \alpha = \frac{3\pi}{8}, \quad \gamma = \frac{\pi}{8}, \quad \beta_A \approx 6.07, \quad \beta_B \approx 5.42.
\end{equation}
\begin{figure}[!htbp]
    \centering
    \includegraphics[width=0.58\linewidth]{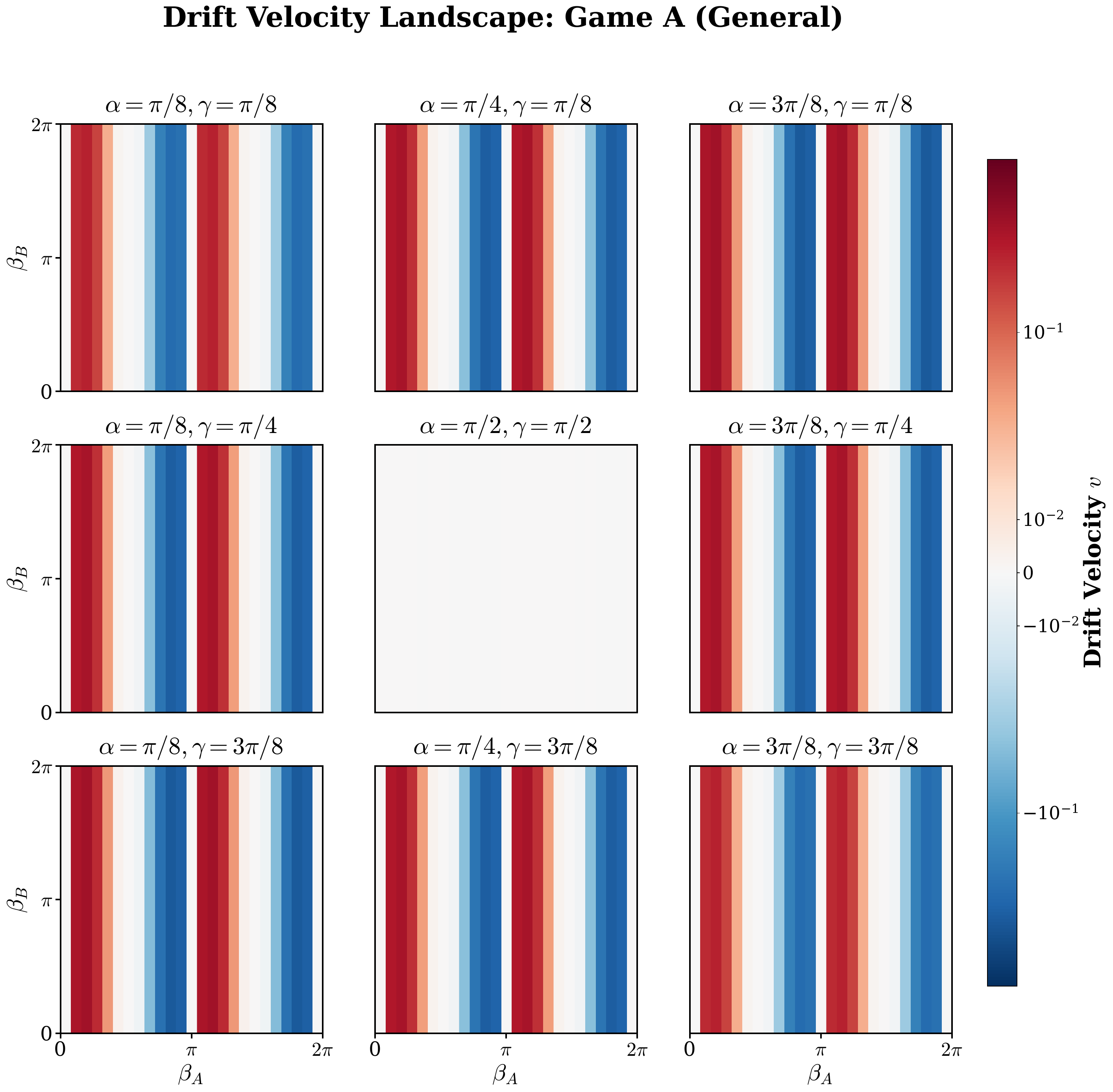}
    \caption{\textbf{Drift velocity landscape for Game~A (General scan).} Vertical stripes indicate dependence solely on $\beta_A$.}
    \label{fig:scan_A}
\end{figure}

\begin{figure}[!htbp]
    \centering
    \includegraphics[width=0.58\linewidth]{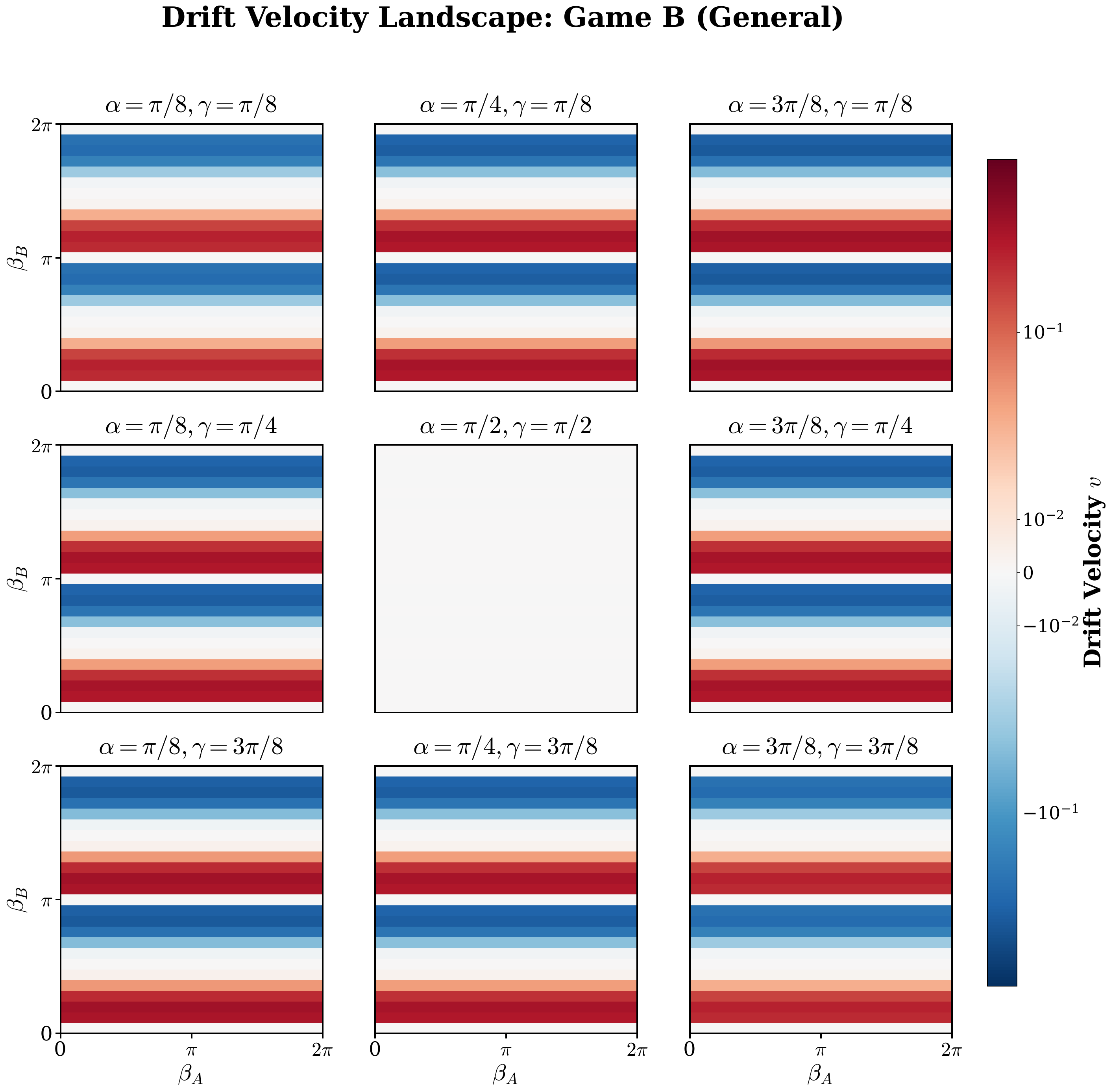}
    \caption{\textbf{Drift velocity landscape for Game~B (General scan).} Horizontal stripes indicate dependence solely on $\beta_B$.}
    \label{fig:scan_B}
\end{figure}

\clearpage
\subsection{Standard Parameter Scan (Discrete Angles)}
\label{sec:standard_scan}
To verify robustness and connect with standard models, we restricted the scan to $\alpha, \gamma \in \{0, \pi/2, \pi\}$. Although the paradoxical regions are narrower, a distinct solution was identified at $\alpha=0$ and $\gamma=\pi/2$. The corresponding best-performing rotation angles are:
\begin{equation}
    \alpha = 0, \quad \gamma = \frac{\pi}{2}, \quad \beta_C \approx 6.12, \quad \beta_D \approx 2.26.
\end{equation}

\begin{figure}[!htbp]
    \centering
    \includegraphics[width=0.58\linewidth]{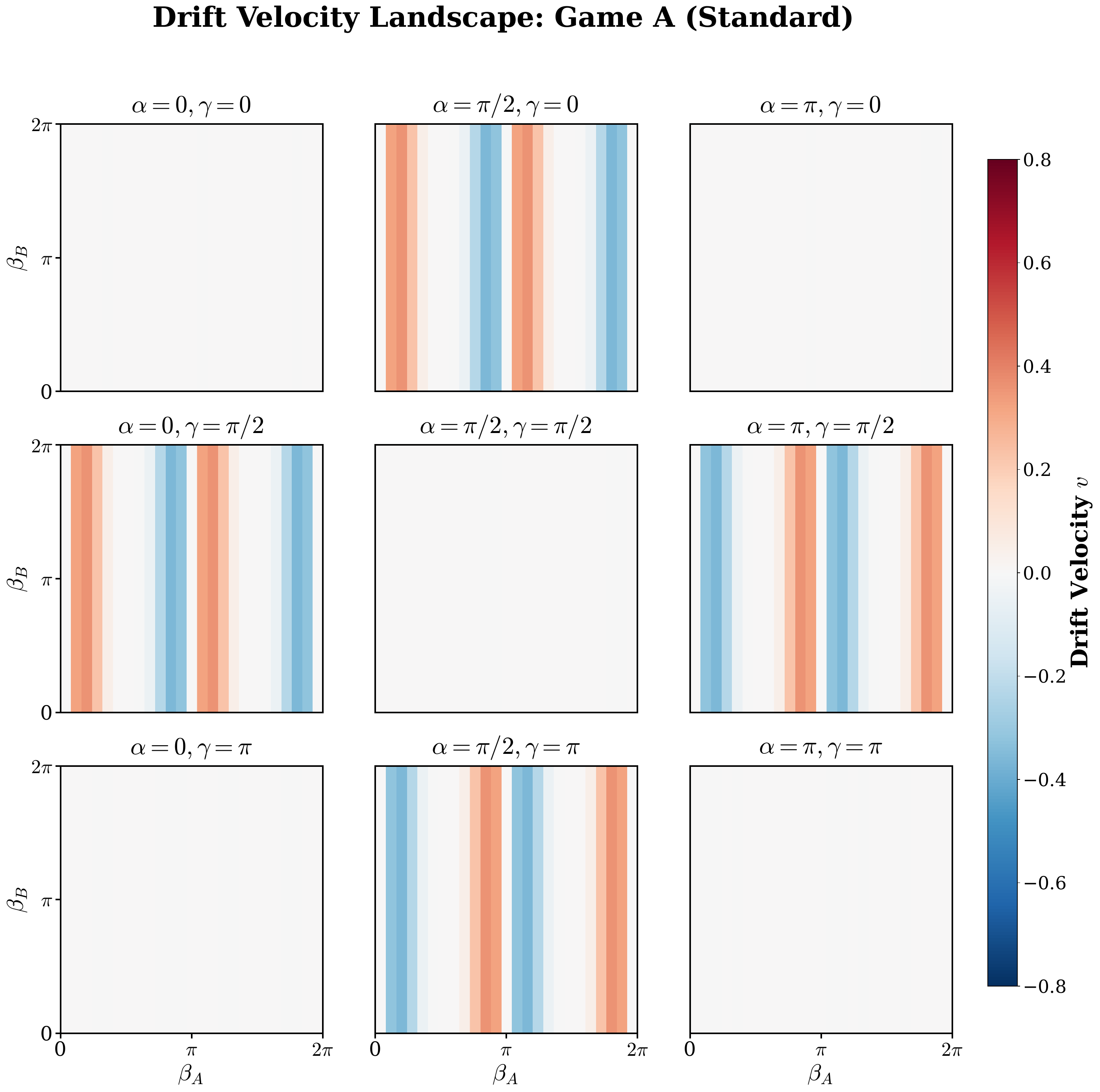}
    \caption{\textbf{Drift velocity landscape for Game~C (Standard scan).} Baseline drift map for standard discrete angles.}
    \label{fig:scan_A_std}
\end{figure}

\begin{figure}[!htbp]
    \centering
    \includegraphics[width=0.58\linewidth]{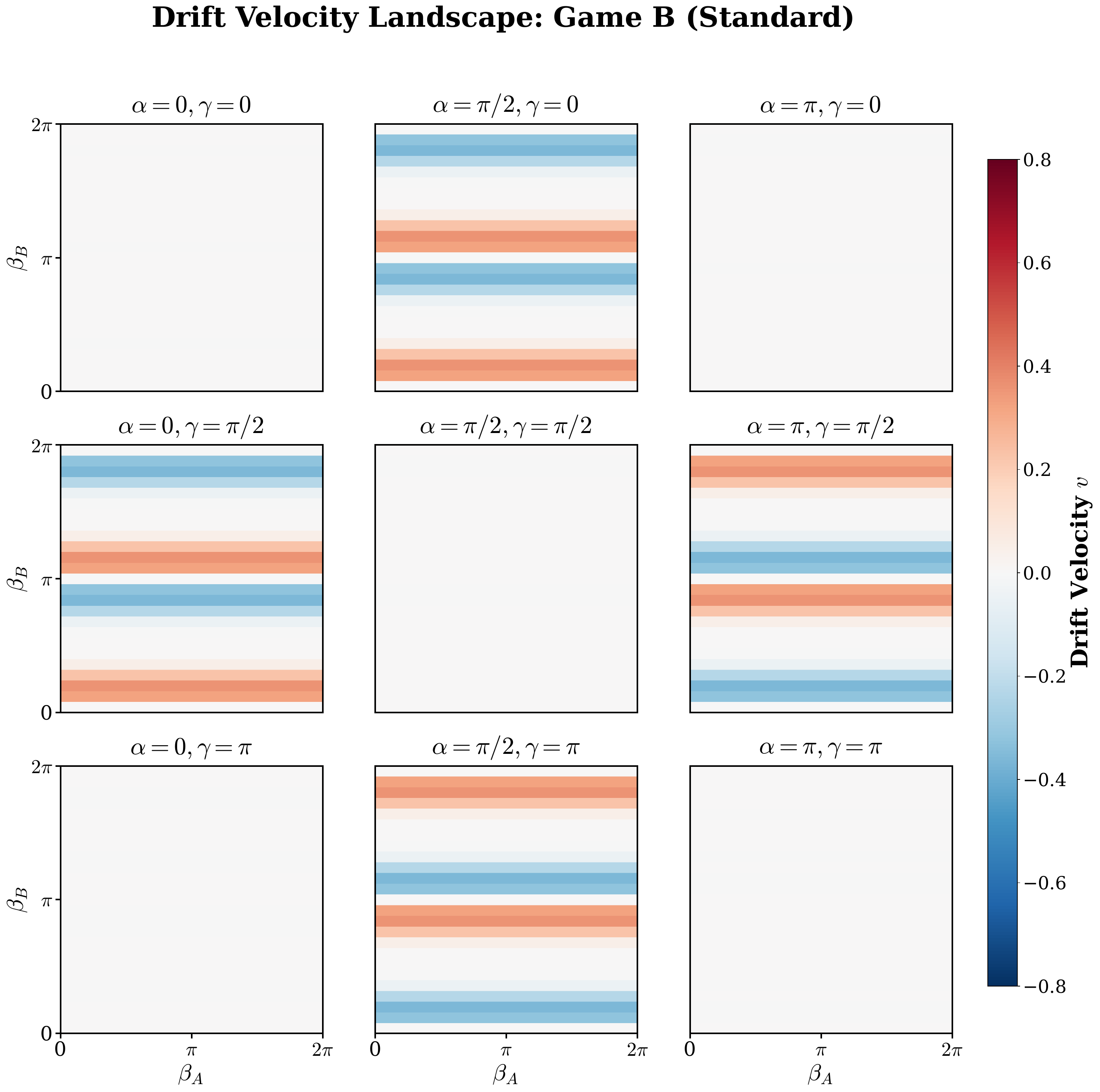}
    \caption{\textbf{Drift velocity landscape for Game~D (Standard scan).} Baseline drift map for standard discrete angles.}
    \label{fig:scan_B_std}
\end{figure}

\begin{figure}[!htbp]
    \centering
    \includegraphics[width=0.58\linewidth]{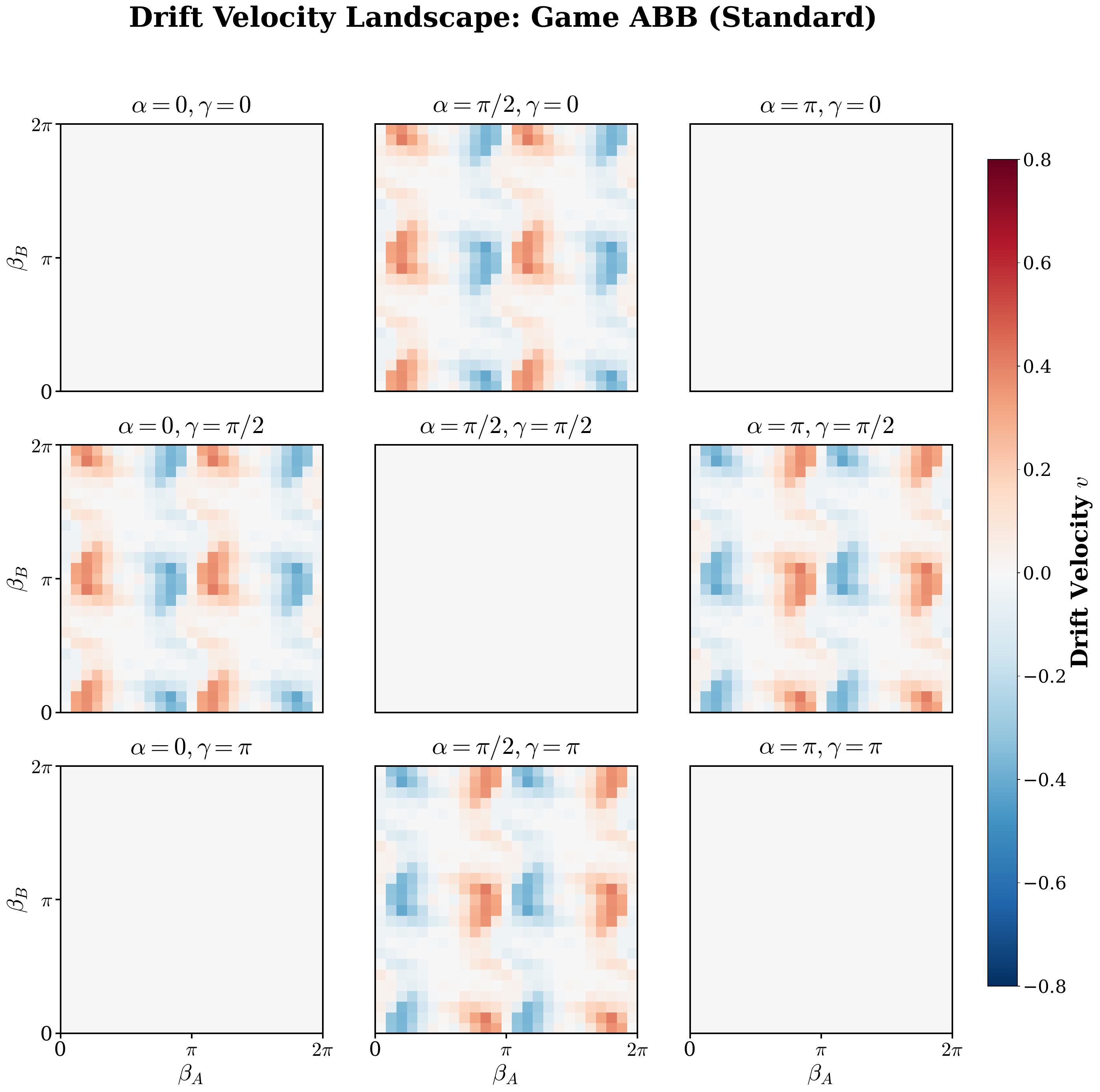}
    \caption{\textbf{Drift velocity landscape for Sequence CDD (Standard).} A clear reversal is observed around $\alpha=0, \gamma=\pi/2$, confirming the paradox persists under standard parameter constraints.}
    \label{fig:scan_ABB_std}
\end{figure}
\end{document}